\documentclass[12pt,preprint]{aastex}
\usepackage{graphicx}
\shortauthors{Milone et al.}
\usepackage{ulem}

\begin{document}
\title{The {\it Hubble Space Telescope} UV Legacy Survey of Galactic Globular Clusters. III. A quintuple stellar population in NGC\,2808.
          \footnote{           Based on observations with  the
                               NASA/ESA {\it Hubble Space Telescope},
                               obtained at  the Space Telescope Science
                               Institute,  which is operated by AURA, Inc.,
                               under NASA contract NAS 5-26555.}}

\author{ A.\,P.\,Milone\altaffilmark{2},
 A.\,F.\,Marino\altaffilmark{2},
 G.\,Piotto\altaffilmark{3,4},
 A.\,Renzini\altaffilmark{4},
 L.\,R.\,Bedin\altaffilmark{4},
 J.\,Anderson\altaffilmark{5},
 S.\,Cassisi\altaffilmark{6},
 F.\,D'Antona\altaffilmark{7},
 A.\,Bellini\altaffilmark{5},
 H.\,Jerjen\altaffilmark{2},
 A.\,Pietrinferni\altaffilmark{6},
 P.\,Ventura\altaffilmark{7}
}

\altaffiltext{2}{Research School of Astronomy and Astrophysics, The Australian National University, Cotter Road, Weston, ACT, 2611, Australia}
\altaffiltext{3}{Istituto Nazionale di Astrofisica - Osservatorio Astronomico di Padova, Vicolo dell'Osservatorio 5, Padova, IT-35122}
\altaffiltext{4}{Dipartimento di Fisica e Astronomia ``Galileo Galilei'', Univ. di Padova, Vicolo dell'Osservatorio 3, Padova, IT-35122}
\altaffiltext{5}{Space Telescope Science Institute, 3800 San Martin Drive, Baltimore,  MD 21218, USA}
\altaffiltext{6}{Istituto Nazionale di Astrofisica - Osservatorio Astronomico di Teramo, Via Mentore  Maggini s.n.c., I-64100 Teramo, Italy}
\altaffiltext{7} {Istituto Nazionale di Astrofisica - Osservatorio Astronomico di Roma, Via Frascati 33, I-00040 Monteporzio Catone, Roma, Italy}

\begin{abstract}
 In this study we present first results from multi-wavelength {\it Hubble Space Telescope} ({\it HST}\/) observations of the Galactic globular cluster (GC) NGC\,2808
as an extension of the {\it Hubble Space Telescope UV Legacy Survey of Galactic GCs} (GO-13297 and previous proprietary and {\it  HST} archive data).
 Our analysis allowed us to disclose a multiple-stellar-population phenomenon in\,NGC\,2808 even more complex than previously thought.
 We  have separated at least five different populations along the main sequence and the red giant branch (RGB), that we name A, B, C, D and E  (though an even finer subdivision may be suggested by the data). 
 We identified the RGB bump in four out of the five RGBs.
 To explore the origin of this complex CMD, we have combined our multi-wavelength {\it HST} photometry with synthetic spectra, generated by assuming different chemical compositions.
The comparison of observed colors with synthetic spectra suggests that the five stellar populations  have different contents of light elements and helium. Specifically, if we assume that NGC\,2808 is homogeneous in [Fe/H] (as suggested by spectroscopy for Populations B, C, D, E, but lacking for Population A)
and that population A has a primordial helium abundance, we find that populations B, C, D, E are enhanced in helium by $\Delta$ Y$\sim$0.03, 0.03, 0.08, 0.13, respectively. We obtain similar results by comparing the magnitude of the RGB bumps with models. 
Planned spectroscopic observations will test whether  also Population A has the same metallicity, or whether its photometric differences with Population B can be ascribed to small [Fe/H] and [O/H] differences rather than to helium.
\end{abstract}


\shorttitle{Multiple populations in NGC\,2808} 
\shortauthors{Milone et al.} 

\section{Introduction}
\label{sec:introduction}
Recent studies, based on multi-wavelength photometry, have revealed that the color-magnitude diagram (CMD) of all Galactic globular clusters (GCs) so far explored (Piotto et al.\ 2015, hereafter Paper\,I) is made of distinct sequences of stars that can be traced continuously from the bottom of the main sequence (MS) up to the tip of the red giant branch (RGB) and through the horizontal branch (HB) and the asymptotic giant branch (AGB). These sequences stand in contrast to the traditional view  of GCs as  the best example of simple stellar populations,  i.e.\,made of stars born all at the same time and with the same chemical composition, confirming previous findings of CN variations in MS and RGB stars (Cannon et al.\,1998; Grundahl et al.\,1998; Grundahl\,1999). \\
 The {\it Hubble Space Telescope UV Legacy Survey of Galactic GCs} is an {\it Hubble Space Telescope} ({\it HST}) project to observe 54 GCs through the filters F275W, F336W, F438W of the Wide Field Camera 3 (WFC3)  onboard {\it HST} (see Paper\,I). This dataset complements the existing F606W and F814W photometry from the Advanced Camera for Survey (GO-10775, Sarajedini et al.\,2007; Anderson et al.\,2008) and is specifically designed to map multiple stellar populations in GCs.

 NGC\,2808 is one of the most intriguing Galactic GCs in the context of multiple stellar populations. It hosts a multimodal MS (D'Antona et al.\,2005; Piotto et al.\,2007; Milone et al.\,2012a; and Paper\,I), and exhibits a multimodal HB (Sosin et al.\,1997; Bedin et al.\,2000; Dalessandro et al.\,2011) and RGB (Lee et al.\,2009;  Monelli et al.\,2013; Paper\,I).
Spectroscopy has shown star-to-star variations of several light elements, lithium and an extended Na-O anticorrelation (Norris \& Smith\,1983; Carretta et al.\,2006, 2010; Gratton et al.\,2011; Marino et al.\,2014; Carretta\,2014; D'Orazi et al.\,2015).
These observations have been interpreted with multiple populations of stars with different helium abundance, from primordial abundance, $Y \sim$0.246, up to extreme enhancement, $Y \sim$0.38 (e.g.\,D'Antona et al.\,2002, 2005; Piotto et al.\,2007; Milone et al.\,2012a). Evidence of helium enhancement in NGC\,2808 has been also confirmed by direct measurements of helium-rich stars along the RGB and the HB (Pasquini et al.\,2011; Marino et al.\,2014).

While previous studies on multiple MSs in NGC\,2808 were based on visual or near-infrared photometry, in this paper we extend the study to the ultraviolet.  The ultraviolet region of the spectrum is indeed very powerful in the study of multiple stellar populations with different chemical composition. Molecular bands, such as OH, NH, CH, and CN affect the ultraviolet and blue wavelengths, that are thus sensitive to populations with different C, N, and O compositions (Milone et al.\,2012b; Paper\,I). 

In this paper we use multi-wavelength ultraviolet and visual photometry (from Paper\,I ) of stars in a field centered on NGC\,2808 in order to identify multiple stellar populations in the CMDs. The behavior of multiple sequences in appropriate CMDs made with different combinations of colors and magnitudes will provide unique information on the helium and light-element content of the different stellar populations of this extreme GC.

The paper is organized as follows.
 In Sect.~\ref{sect:data} we present the data and the data reduction. In Sect.~\ref{sec:multiseq} we analyze the CMDs and investigate multiple populations along the RGB and the MS.  Helium and C, N, O abundances of the stellar populations are inferred in Sect.~\ref{helium} from multiple MS and RGB
 locations in the CMD. The bump of multiple RGBs of NGC\,2808 are analyzed in Sect.~\ref{sec:bump}, while Sects.~\ref{sec:HB} and ~\ref{sec:agb} are dedicated to the AGB and the HB. A discussion will follow in  Sect.~\ref{sec:discussion}.

\section{Data and Data Reduction}
\label{sect:data}
In our study of NGC\,2808 we have used archival and proprietary images taken with the Wide Field Channel of the Advanced Camera for Surveys (WFC/ACS) and the Ultraviolet and Visual Channel of the Wide Field Camera 3 (UVIS/WFC3) on board of the {\it Hubble Space Telescope} ({\it HST}).
Table~1 summarizes the dataset. 

The poor charge-transfer efficiency (CTE) in the UVIS/WFC3 and ACS/WFC images have been corrected by following the recipe by Anderson \& Bedin\,(2010). 
Photometry and astrometry of UVIS/WFC3 images were already presented in Paper\,I and were obtained with $img2xym\_UVIS\_09\times10$, which is a software package presented by Bellini et al.\,(2010) and mostly adapted from $img2xym\_WFI$ (Anderson et al.\,2006). We used pixel-area and geometric-distortion corrections from Bellini \& Bedin (2009) and Bellini, Anderson \& Bedin (2011). The photometry has been calibrated as in Bedin et al.\,(2005), and is using the encircled energy and the zero points available at the STScI web page.
We used the photometric and astrometric catalogs from WFC/ACS data published by Anderson et al.\,(2008), Sarajedini et al.\,(2007), Milone et al.\,(2012a), Piotto et al.\,(2007), which were obtained from GO-9899, GO-10922, and GO-10775 WFC/ACS data.

In order to investigate multiple stellar populations in NGC\,2808 we are interested in stars for which high-accuracy photometry is available.  The stellar catalogs were purged of stars that are poorly measured by using the procedure described by Milone et al.\,(2009) and based on the quality indexes provided by our software (see Anderson et al.\,2006, 2008). Photometry has been corrected for differential reddening following the recipe in Milone et al.\,(2012a).

\begin{table}[!htp]
\center
\scriptsize {
\begin{tabular}{cccccl}
\hline
\hline
 INSTR.\, &  DATE & N$\times$EXPTIME & FILTER  & PROGRAM & PI \\
\hline
ACS/WFC & May 5 2004 & 6$\times$340s & F475W & 9899 & G.\,Piotto\\
ACS/WFC & Aug 9 and Nov 2 2006 & 20s$+$2$\times$350s$+$2$\times$360s & F475W & 10922 & G.\,Piotto\\
ACS/WFC & Aug 9 and Nov 1 2006 & 10s$+$3$\times$350s$+$3$\times$360s & F814W & 10922 & G.\,Piotto\\
ACS/WFC & Jan 1 2006 & 23s$+$5$\times$360s & F606W & 10775 & A.\,Sarajedini\\
ACS/WFC & Jan 1 2006 & 23s$+$5$\times$370s & F814W & 10775 & A.\,Sarajedini\\
WFC3/UVIS & Sep 08-09 2013 &12$\times$985s & F275W & 12605 &  G.\,Piotto \\
WFC3/UVIS & Sep 08-09 2013 & 6$\times$650s & F336W & 12605 &  G.\,Piotto \\
WFC3/UVIS & Sep 08-09 2013 & 6$\times$97s  & F438W & 12605 &  G.\,Piotto \\
\hline
\hline
\end{tabular}
}
\label{tab:data}
\caption{List of the data sets used in this paper. }
\end{table}

\section{The multiple photometric components along the CMD of NGC\,2808}
\label{sec:multiseq}
As already discussed in Sect.~\ref{sec:introduction}, 
previous studies based on ACS/{\it HST} and ground-based photometry
 have shown that NGC\,2808 has at least a triple MS (Piotto et al.\,2007; Milone et al.\,2012a) and a broadened RGB (Lee et al.\,2009; Monelli et al.\,2013; Paper\,I). 
An inspection of the large number of CMDs that we derived from six-band photometry immediately reveals that NGC\,2808 is even more complex than initially thought, and hosts more than three stellar populations.

\begin{figure*}[htp!]
   \centering
   \epsscale{.99}
      \plotone{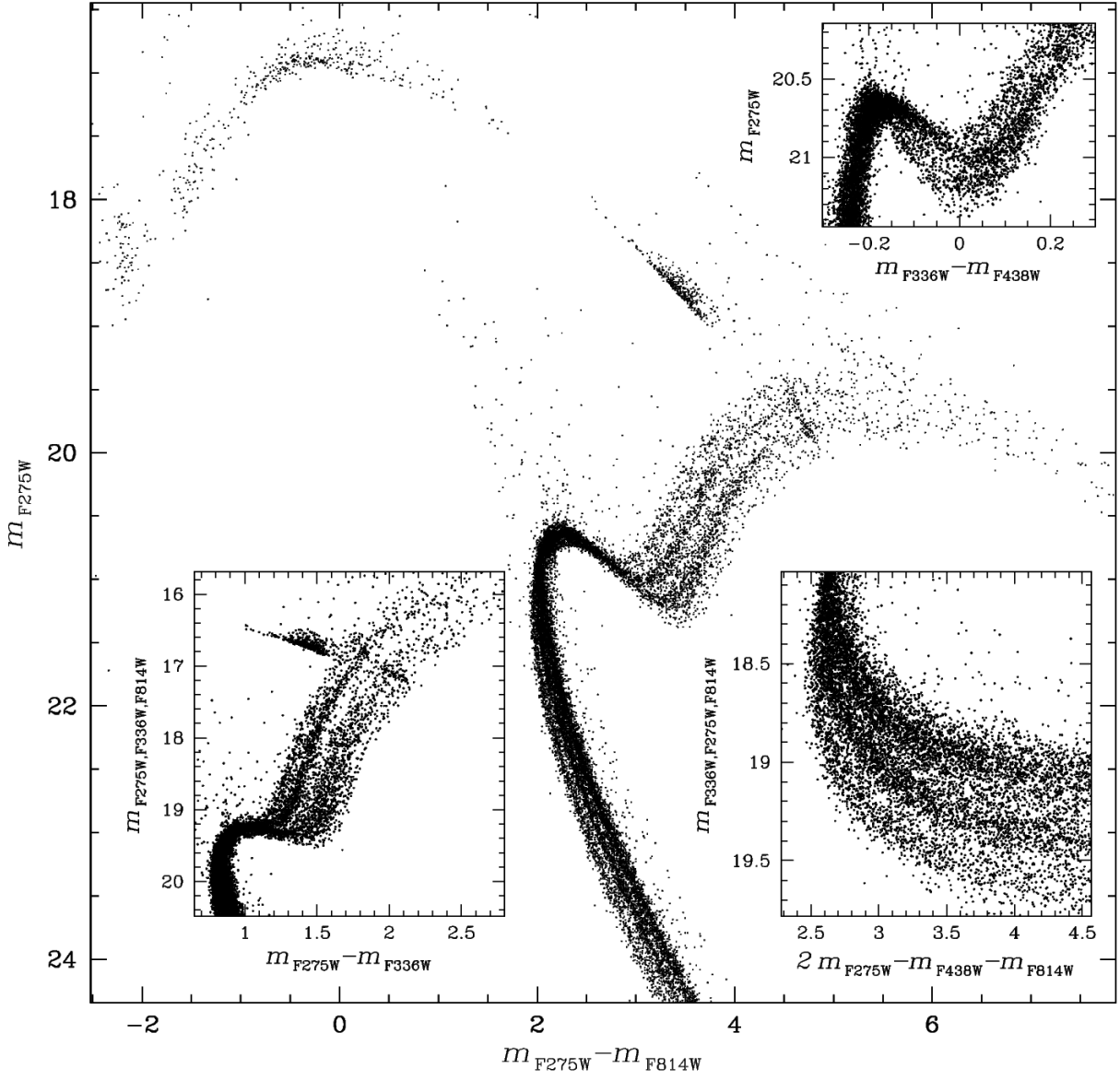}
      \caption{$m_{\rm F275W}$ vs.\,$m_{\rm F275W}-m_{\rm F814W}$ CMD of NGC\,2808. 
      The $m_{\rm F275W,F336W,F814W}$ against $m_{\rm F275W}-m_{\rm F336W}$ (bottom-left inset), $m_{\rm F336W,F275W,F814W}$ against $2~m_{\rm F275W}-m_{\rm F438W}-m_{\rm F814W}$ (bottom-right inset), and $m_{\rm F275W}$ vs.\,$m_{\rm F336W}-m_{\rm F438W}$ (upper-right inset) diagrams highlight multiple sequences along the RGB, the MS, and the SGB, respectively.}
          \label{SummaryNGC2808}
   \end{figure*}

 A visual example of its complexity is provided by the CMDs in Fig.~\ref{SummaryNGC2808}. This figure  shows several diagrams, after the quality selection and the differential-reddening correction described in the previous section were applied. 
To derive some  of the diagrams of Fig.~\ref{SummaryNGC2808} we have defined the pseudo-magnitudes $m_{\rm F336W,F275W,F814W}=(m_{\rm F336W}-m_{\rm F275W}+m_{\rm F814W})$ and $m_{\rm F275W,F336W,F814W}=(m_{\rm F275W}-m_{\rm F336W}+m_{\rm F814W})$, which allow us to better distinguish multiple sequences along the RGB and the MS.
 An inspection of these CMDs immediately suggests that
both the RGB and MS are made of multiple sequences, that look discrete in the $m_{\rm F275W}$ vs.\,$m_{\rm F275W}-m_{\rm F814W}$, $m_{\rm F275W,F336W,F814W}$ vs.\,$m_{\rm F275W}-m_{\rm F336W}$, and $m_{\rm F336W,F275W,F814W}$ vs.\,$2~m_{\rm F275W}-m_{\rm F438W}-m_{\rm F814W}$ diagrams. 
We also observe a widely-spread SGB in the $m_{\rm F275W}$ vs.\,$m_{\rm F336W}-m_{\rm F438W}$ CMD as shown in the upper-right inset of Fig.~\ref{SummaryNGC2808}. 
 In the following we discuss the observed morphology of the CMD at various evolutionary stages: the RGB, MS, AGB, and HB.

\subsection{The quintuple RGB}
\label{sec:RGBs}

\begin{figure*}[htp!]
   \centering
   \epsscale{.99}
      \plotone{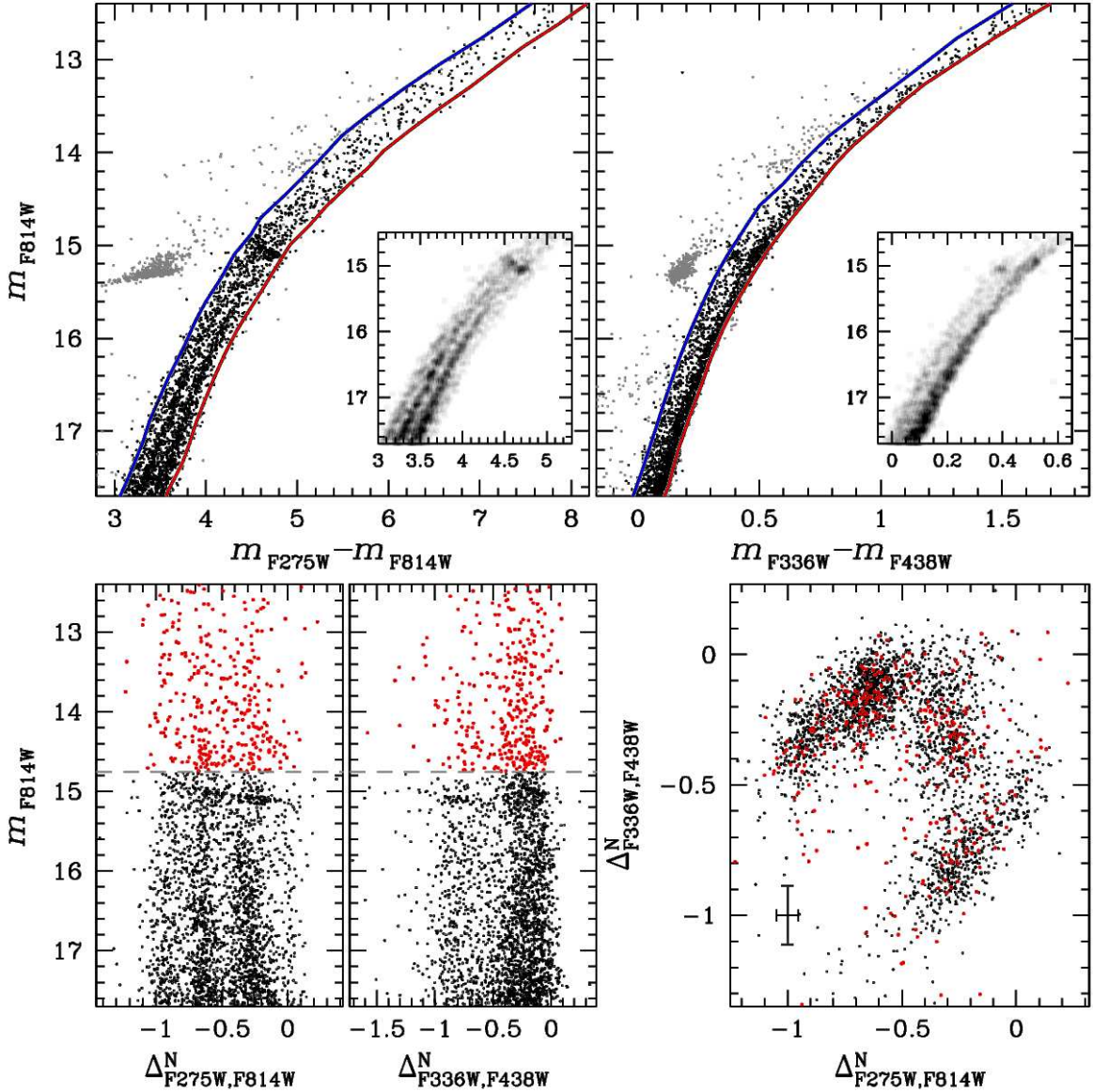}
      \caption{\textit{Upper panels:} Zoom of the $m_{\rm F814W}$ vs.\,$m_{\rm F275W}-m_{\rm F814W}$ (left) and of the $m_{\rm F814W}$ vs.\,$m_{\rm F336W}-m_{\rm F438W}$                       (right) CMD of NGC\,2808 around the RGB. Only RGB stars colored black are used in the following analysis. Red and blue lines are the fiducials adopted to verticalize the RGB (see text for details).
 The insets show the Hess diagram for RGB stars with $12.25<m_{\rm F814W}<17.7$.
    \textit{Lower panels:} Verticalized $m_{\rm F814W}$ vs.\,$\Delta^{\rm N}_{\rm F275W, F814W}$ (left) and $m_{\rm F814W}$ vs.\,$\Delta^{\rm N}_{\rm F336W, F438W}$ (middle) diagrams for RGB stars. $\Delta^{\rm N}_{\rm F336W, F438W}$ is plotted against $\Delta^{\rm N}_{\rm F275W, F814W}$ in the lower-right panel.
  RGB stars with $m_{\rm F814W}<14.68$  are colored red.
      }
          \label{NGC2808pops}
   \end{figure*}

 Along the RGB, the behavior of multiple populations dramatically changes from one CMD to another. In order to investigate this phenomenon, in Fig.~\ref{NGC2808pops} we compare the $m_{\rm F814W}$ vs.\,$m_{\rm F275W}-m_{\rm F814W}$ CMD of RGB stars (upper-left panel) and of the $m_{\rm F814W}$ vs.\,$m_{\rm F336W}-m_{\rm F438W}$ CMD (upper-right panel). The insets show the Hess diagrams for stars in the magnitude interval with $14.5<m_{\rm F814W}<17.7$ where multiple RGBs are clearly visible. Multiple stellar populations manifest themselves as four separate sequences in $m_{\rm F275W}-m_{\rm F814W}$, while the $m_{\rm F336W}-m_{\rm F438W}$ color distribution is more broadened and only two RGBs can be recognized.

In order to compare the two CMDs, we used the procedure introduced by Milone et al.\,(2015, hereafter paper\,II) in their study of multiple populations in M\,2, and illustrated in Fig.~\ref{NGC2808pops} for the case of NGC\,2808. For that purpose, we drew two fiducial lines in each CMD.
The blue and the red fiducial mark the bluest and the reddest envelope of the RGB, respectively and have been derived as follows.
 We have divided the RGB portion with $m_{\rm F814W}>$14.8 into intervals of 0.2 magnitudes in F814W band. For each interval, we have determined the $4^{th}$ and the $96^{th}$ percentile of the $m_{\rm F275W}-m_{\rm F814W}$ or $m_{\rm F336W}-m_{\rm F438W}$ color distribution and the median $m_{\rm F814W}$ magnitude. The points corresponding to the $4^{th}$ percentile and the median magnitude have been interpolated with a cubic spline to derive the blue fiducial, while the red fiducial has been similarly derived.
Due to small number statistics it is not possible to infer robust estimates of the RGB envelopes with this method at brighter luminosities.
 Therefore, the portions of the blue and the red line in the magnitude interval with $m_{\rm F814W}<$14.8 have been derived by hand trying to follow the blue and the red envelope of the RGB, respectively.

Then we have verticalized the two CMDs in a way that the blue and the red fiducials translate into vertical lines with abscissa $-$1 and 0, respectively. 
To do this, we defined for each star:\\ 
$\Delta_{\rm X}^{\rm N}$=[($X-X_{\rm blue~fiducial}$)/($X_{\rm red~fiducial}-X_{\rm blue~fiducial}$)]-1 where $X$=($m_{\rm F275W}-m_{\rm F814W}$), ($m_{\rm F336W}-m_{\rm F438W}$) and $X_{\rm blue~fiducial}$ and $X_{\rm red~fiducial}$ are obtained by subtracting the color of the fiducial at the corresponding F814W magnitude from the color of each star. The verticalized $m_{\rm F814W}$ vs.\,$\Delta^{\rm N}_{\rm F275W, F814W}$ and $m_{\rm F814W}$ vs.\,$\Delta^{\rm N}_{\rm F336W, F438W}$ diagrams are plotted in the lower-left panels of Fig.~\ref{NGC2808pops}.  RGB stars in NGC\,2808 are clustered around distinct values of $\Delta^{\rm N}_{\rm F336W, F438W}$ and $\Delta^{\rm N}_{\rm F275W, F814W}$, as shown in the bottom-right panel of Fig.~\ref{NGC2808pops}.

As previously discussed by Anderson et al.\,(2008, see their Sect.~8.1), F814W photometry of bright RGB stars is less accurate than that of the remaining RGB stars because it has been derived by using saturated stars (see Anderson et al.\,2008 for details). Indeed, 
multiple sequences are less evident above the gray dashed lines in the lower-left panels of Fig.~\ref{NGC2808pops}. Dashed lines are placed at $m_{\rm F814W}=14.68$. To investigate whether the distinct sequences can be  also detected
along the brightest RGB segment or not, we have marked   stars with $m_{\rm F814W}<14.68$ with red dots in the  lower panels of Fig.~\ref{NGC2808pops}. 
The distribution of these bright RGB stars on the $\Delta^{\rm N}_{\rm F336W, F438W}$ vs.\,$\Delta^{\rm N}_{\rm F275W, F814W}$ plot shows that these stars share the same color distribution as the fainter RGB stars.
%

\begin{figure*}[htp!]
   \centering
   \epsscale{.75}
      \plotone{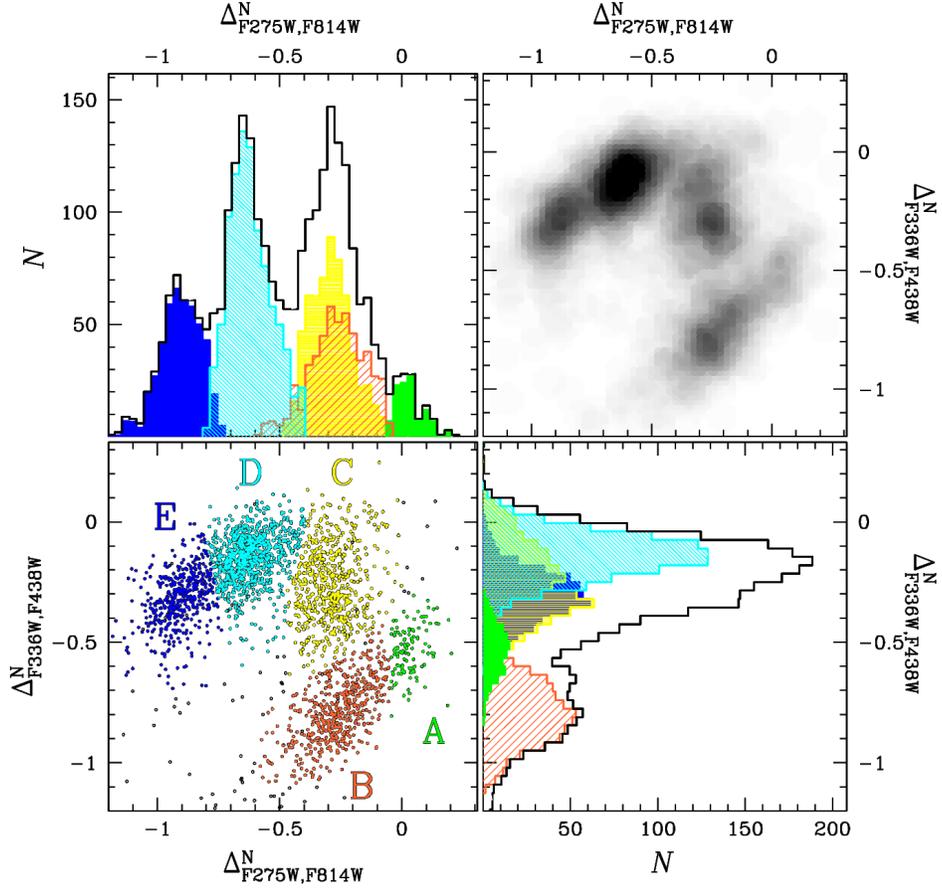}
      \caption{ Reproduction of the $\Delta_{\rm F336W, F438W}$ vs.\,$\Delta_{\rm F275W, F814W}$ diagram of Fig.~\ref{NGC2808pops}. Stars in the A, B, C, D, and E group are colored green, orange, yellow, cyan, and blue, respectively (lower-left).   The corresponding Hess diagram is plotted in the upper-right panel.  The histograms of the normalized $\Delta_{\rm F275W, F814W}$ and $\Delta_{\rm F336W, F438W}$ distributions for all the analyzed RGB stars are plotted in black in the upper-left and lower-right panel, respectively. The shaded colored histograms show the distributions for each of the five populations defined in the lower-left panel.}
          \label{seleRGBs}
   \end{figure*}

To further investigate the stellar populations along the RGB, in the upper-right panel of Fig.~\ref{seleRGBs} we plot the $\Delta^{\rm N}_{\rm F336W, F438W}$ vs.\,$\Delta^{\rm N}_{\rm F275W, F814W}$ Hess diagram. At least five main clumps of RGB stars are clearly visible. These are selected by eye and designated A, B, C, D, and E and are  colored green, orange, yellow, cyan, and blue, respectively (see the lower-left panel diagram). These color codes will be consistently used in the paper. RGB-A--E contain 5.8$\pm$0.5\%, 17.4$\pm$0.9\%, 26.4$\pm$1.2\%, 31.3$\pm$1.3\%, and 19.1$\pm$1.0\% of the total number of RGB stars with $12.25<m_{\rm F814W}<17.70$, respectively.  In Section~\ref{sub:chem} we show that populations A--E have different chemical composition.

The $\Delta^{\rm N}_{\rm F275W, F814W}$ and $\Delta^{\rm N}_{\rm F336W, F438W}$ distributions of RGB stars are shown in the upper-left and lower-right panel of Fig.~\ref{seleRGBs}, respectively. 
Black histograms represent the whole sample of RGB stars shown in the lower-left panel, while the distributions of the five distinct RGBs are plotted with shaded-colored histograms.
The $\Delta^{\rm N}_{\rm F275W, F814W}$ and $\Delta^{\rm N}_{\rm F336W, F438W}$ distributions exhibit significant differences.
The histogram distribution of $\Delta_{\rm F275W, F814W}$ clearly shows three main peaks at $\Delta^{\rm N}_{\rm F275W, F814W} \sim$ $-$0.9,$-$0.6, and $-$0.3. 
 The first and the second clumps are mainly composed of population-E and population-D stars, respectively, while the third peak is a mix of both population-B and population-C stars.  
A less populous peak, corresponding to population A, is located at $\Delta^{\rm N}_{\rm F275W, F814W} \sim$0.0. In contrast, the $\Delta^{\rm N}_{\rm F336W, F438W}$ distribution looks bimodal. Most of the stars of populations C, D, and E have $\Delta^{\rm N}_{\rm F336W, F438W}>-0.5$ and determine the main peak at $\Delta^{\rm N}_{\rm F336W, F438W} \sim$ $-$0.2.
 A second peak, mostly composed of population-B stars, is located around $\Delta^{\rm N}_{\rm F336W, F438W} \sim -$0.8.

In addition we note that: 
\begin{itemize}
\item 
Populations B and C are mixed in the $\Delta^{\rm N}_{\rm F275W, F814W}$ color range while they have distinct $\Delta^{\rm N}_{\rm F336W, F438W}$ values, with population-B stars having also smaller $\Delta^{\rm N}_{\rm F336W, F438W}$ values.
\item Population-A stars have larger $\Delta^{\rm N}_{\rm F275W, F814W}$ than both populations B and C. The color order is different in $\Delta_{\rm F336W, F438W}$, where the histogram of population-A stars is located between the histograms of population B and C.   
\item Since, the analyzed RGB stars cover the same F814W magnitude interval and have similar $m_{\rm F336W}$ and $m_{\rm F438W}$ magnitudes, their photometric errors are similar. We note that the $\Delta^{\rm N}_{\rm F336W, F438W}$ spread for populations B and C, ($\sigma_{\rm  \Delta F336W, F438W}^{\rm N, B}=$0.13$\pm$0.01, and $\sigma_{\rm \Delta F336W, F438W}^{\rm N,C}=$0.16$\pm$0.01) are significantly larger than the spread observed for population-A, D and E stars ($\sigma_{\rm \Delta F336W, F438W}^{\rm N,D}=$0.10$\pm$0.02, $\sigma_{\rm \Delta F336W, F438W}^{\rm N,D}=$0.10$\pm$0.01, and $\sigma_{\rm \Delta F336W, F438W}^{\rm N,E}=$0.10$\pm$0.01). This fact indicates that the $m_{\rm F336W}-m_{\rm F438W}$ color spread observed for population B and C is, in part, intrinsic and that both groups B and group C are not simple stellar populations.
 In fact, a visual inspection of the Hess diagram of Fig.~\ref{seleRGBs} suggests that both groups consist of two clumps of stars which are clustered around $\Delta^{\rm N}_{\rm F336W, F438W} \sim -$0.85, $-$0.65 (group B) and $-$0.2, and $-$0.1 (group C)  thus suggesting that stars in both groups B and C do not have homogeneous chemical composition. More data are needed to establish whether these clumps correspond to distinct stellar populations.  
\end{itemize}

  The causes of the `discreteness' of multiple populations as observed in the CMD and two-color diagram of some GCs are still unknown, and have been associated with distinct bursts of star formation (see Renzini\,2008 for a critical discussion). The referee has pointed out that the distinct bumps in the diagrams of Fig.~\ref{seleRGBs} could indicate that some abundances are favored over the others and suggested a possible connection between the abundances of stars in the distinct clumps of NGC\,2808 and metal mixtures that are consistent with equilibrium CN or equilibrium ON cycling. While this hypothesis deserves some investigation that is beyond the purposes of our paper, we emphasize that the evidence of discrete populations in NGC\,2808 provides strong constraint for any model of formation and evolution of stellar populations in GCs.
 
  Upon request of the referee and of the {\it Statistical Editor} of the journal, Prof.\,Eric Feigelson, we have used the Mcluster CRAN package in the public domain R statistical software system to estimate how many groups are statistically significant.  
This package is based on the method described in details in the monograph `Finite Mixture Models' by McLachlan \& Peel (2000). It performs the maximum likelihood fits to different number of stellar groups, and evaluate the number of groups by the Bayesian Information Criterion (BIC) penalized likelihood measure for model complexity. 

To do this it uses several different assumptions about shape and size of the different populations in a plot such as that shown in Fig.~\ref{seleRGBs}. For each shape and size that we adopted for the populations, we assumed a number, N, of stellar populations from 1 to 20 and estimated a BIC for each combination.
We obtain the best BIC value (BIC=1784) for N=6 under the assumption the the stellar populations have equal shapes but variable volume and orientations (VEV). The second most-likely explanation (BIC=1778) corresponds to N=6 but assumes equal shape, volume, and orientation. The third-best value (BIC=1776) corresponds to a VEV assumption and seven stellar populations. All the three best models assume ellipsoidal distributions. 

Results from this statistical analysis support the conclusion that our observations of NGC\,2808 are consistent with more than five groups of stars, and that group C hosts more than one stellar population. The third best BIC value suggests that also the group B is not consistent with a simple population. Thus, the statistical analysis confirms what was already pretty evident from a pure eye inspection of the plots.  
In the following, we will study the five most-evident stellar populations, A--E. 

\subsection{Multiple populations along the MS}
\label{sub:MS}
\begin{figure}[htp!]
   \centering
   \epsscale{.99}
      \plotone{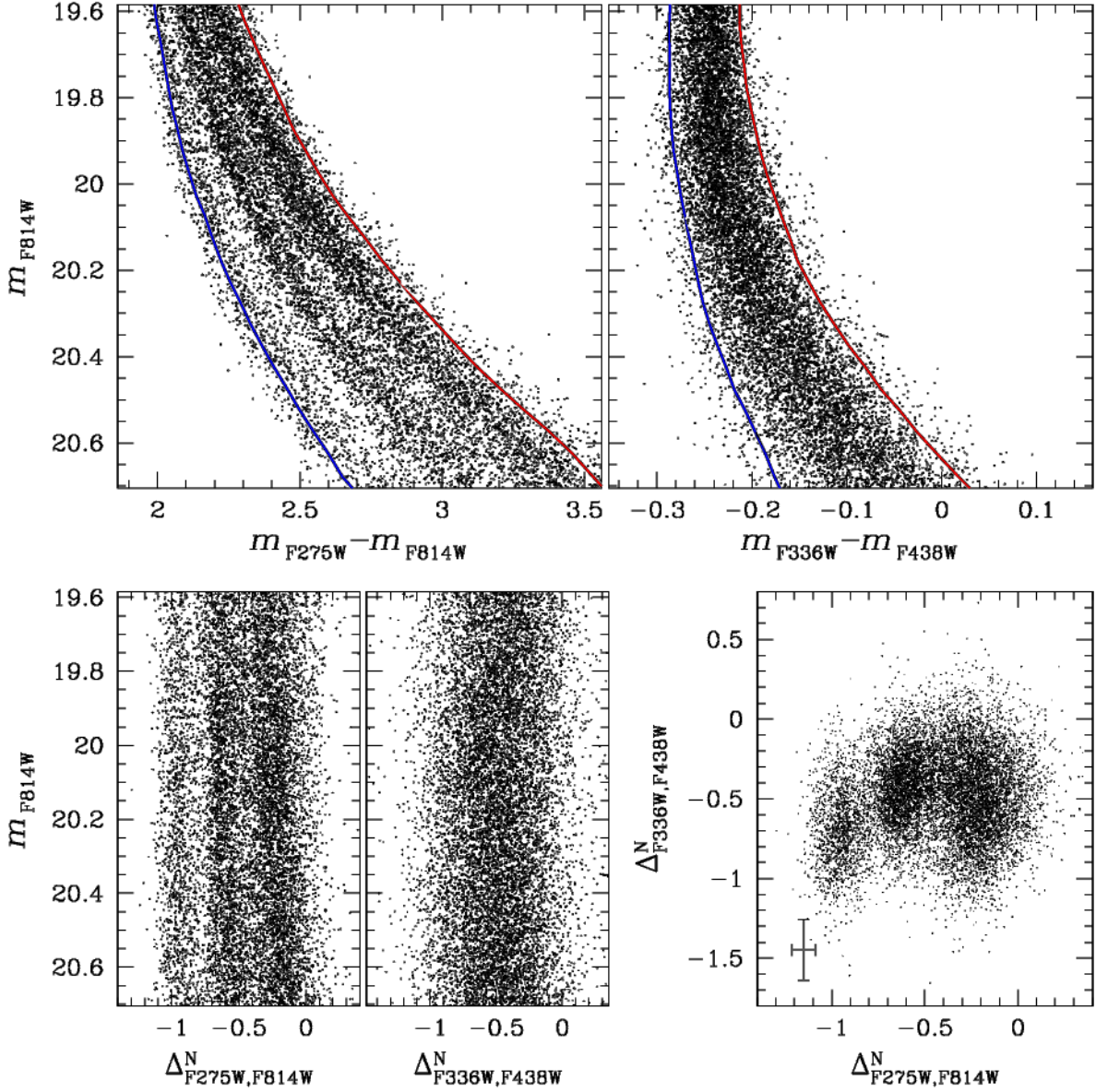}
      \caption{ \textit{Upper panels:} $m_{\rm F814W}$ vs.\,$m_{\rm F275W}-m_{\rm F814W}$ (left) and of the $m_{\rm F814W}$ vs.\,$m_{\rm F336W}-m_{\rm F438W}$ (right)
                        CMD for MS stars in NGC\,2808. 
                        The fiducials used to verticalize the RGB are represented with red and blue lines (see text for details).
    \textit{Lower panels:} Verticalized $m_{\rm F814W}$ vs.\,$\Delta^{\rm N}_{\rm F275W, F814W}$ (left) and $m_{\rm F814W}$ vs.\,$\Delta^{\rm N}_{\rm F336W, F438W}$ (middle) diagram for the stars in the upper panels. $\Delta^{\rm N}_{\rm F336W, F438W}$ is plotted against $\Delta^{\rm N}_{\rm F275W, F814W}$ in the lower-right panel.}
          \label{NGC2808mMSs}
   \end{figure}

The MS of NGC\,2808 exhibits different patterns in CMDs based on different photometric bands,in close analogy with what we observe along the RGB. This is shown in the upper panels of Fig.~\ref{NGC2808mMSs} where we compare the $m_{\rm F814W}$ vs.\,$m_{\rm F275W}-m_{\rm F814W}$ (left panel) and the $m_{\rm F814W}$ vs.\,$m_{\rm F336W}-m_{\rm F438W}$ (right panel) CMDs of MS stars with $19.6<m_{\rm F814W}<20.7$.  The MS looks discrete in $m_{\rm F275W}-m_{\rm F814W}$ with three distinct components, in contrast with the $m_{\rm F336W}-m_{\rm F438W}$ color distribution, which looks broadened without any evidence for discrete sequences.

In order to identify the  different stellar populations, we have verticalized the MSs by following the same recipe as introduced in Sect.~\ref{sec:RGBs} for the RGB, and using the fiducial lines drawn in the upper-panel CMDs. The $m_{\rm F814W}$ vs.\,$\Delta^{\rm N}_{\rm F275W, F814W}$ and the $m_{\rm F814W}$ vs.\,$\Delta^{\rm N}_{\rm F336W, F438W}$  diagrams are plotted in the lower-left and lower-middle panels, while the lower-right panel shows $\Delta^{\rm N}_{\rm F336W, F438W}$ against $\Delta^{\rm N}_{\rm F275W, F814W}$.

\begin{figure}[htp!]
   \centering
   \epsscale{.75}
      \plotone{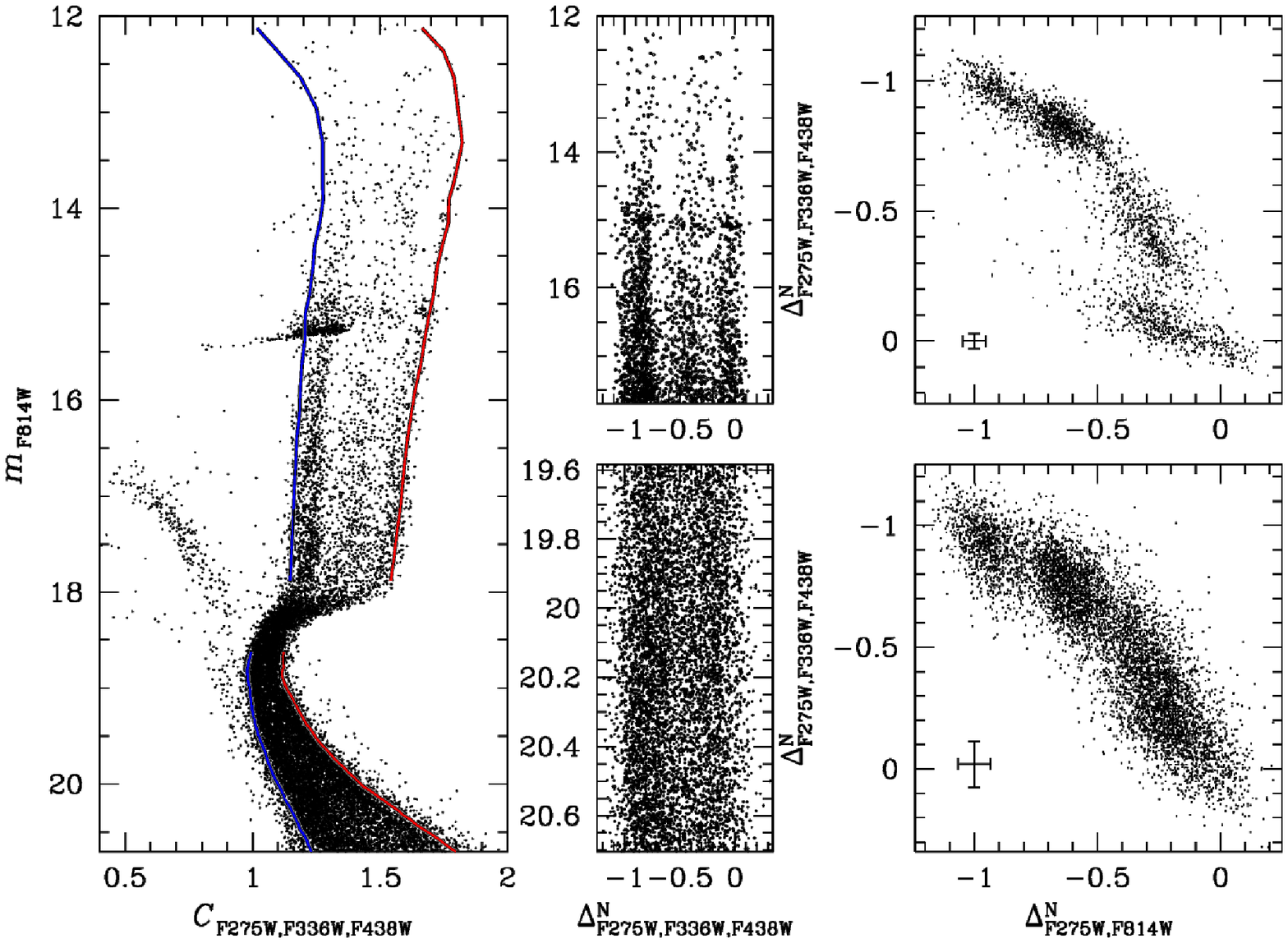}
      \caption{ \textit{Left:} $m_{\rm F814W}$ vs.\,$C_{\rm F275W, F336W, F438W}$ diagram for NGC\,2808. The red and blue lines superimposed on the diagram are the fiducial lines used to verticalize the MS and the RGB. See text for details. \textit{Middle:} Verticalized MS (upper panel) and RGB (lower panel). \textit{Right:}  $\Delta^{\rm N}_{\rm F275W, F336W, F438W}$ vs.\,$\Delta^{\rm N}_{\rm F275W, F814W}$ for the RGB and MS stars plotted in the middle panels.}
          \label{NGC2808mMSs2}
   \end{figure}

The pseudo color $C_{\rm F275W, F336W, F438W}$=($m_{\rm F275W}-m_{\rm F336W}$)$-$($m_{\rm F336W}-m_{\rm F438W}$) defined by Milone et al.\,(2013) is another valuable tool to identify multiple populations in GCs. To better distinguish the distinct MSs and RGBs of NGC\,2808 we show in the left panel of Fig.~\ref{NGC2808mMSs2} the $m_{\rm F814W}$ vs.\,$C_{\rm F275W, F336W, F438W}$ pseudo CMD for this cluster. The red and the blue lines superimposed on this diagram are the envelopes of the MS and the RGB, and have been determined with the same procedure as in Sect.~\ref{sec:RGBs}. These two fiducials are used to verticalize the MS and the RGB.
 The verticalized $m_{\rm F814W}$ vs.\,$\Delta^{\rm N}_{C \rm F275W,F336W,F438W}$ diagram for RGB and MS stars are plotted in the middle panels. An inspection of these figures reveals that three distinct sequences are present along the RGB, while only two MSs are visible.
In the right panels of Fig.~\ref{NGC2808mMSs2} we plot $\Delta^{\rm N}_{\rm F275W, F814W}$ against $\Delta^{\rm N}_{C \rm F275W, F336W, F438W}$ for RGB (upper panel) and MS stars (lower panel).

\begin{figure}[htp!]
   \centering
   \epsscale{.75}
      \plotone{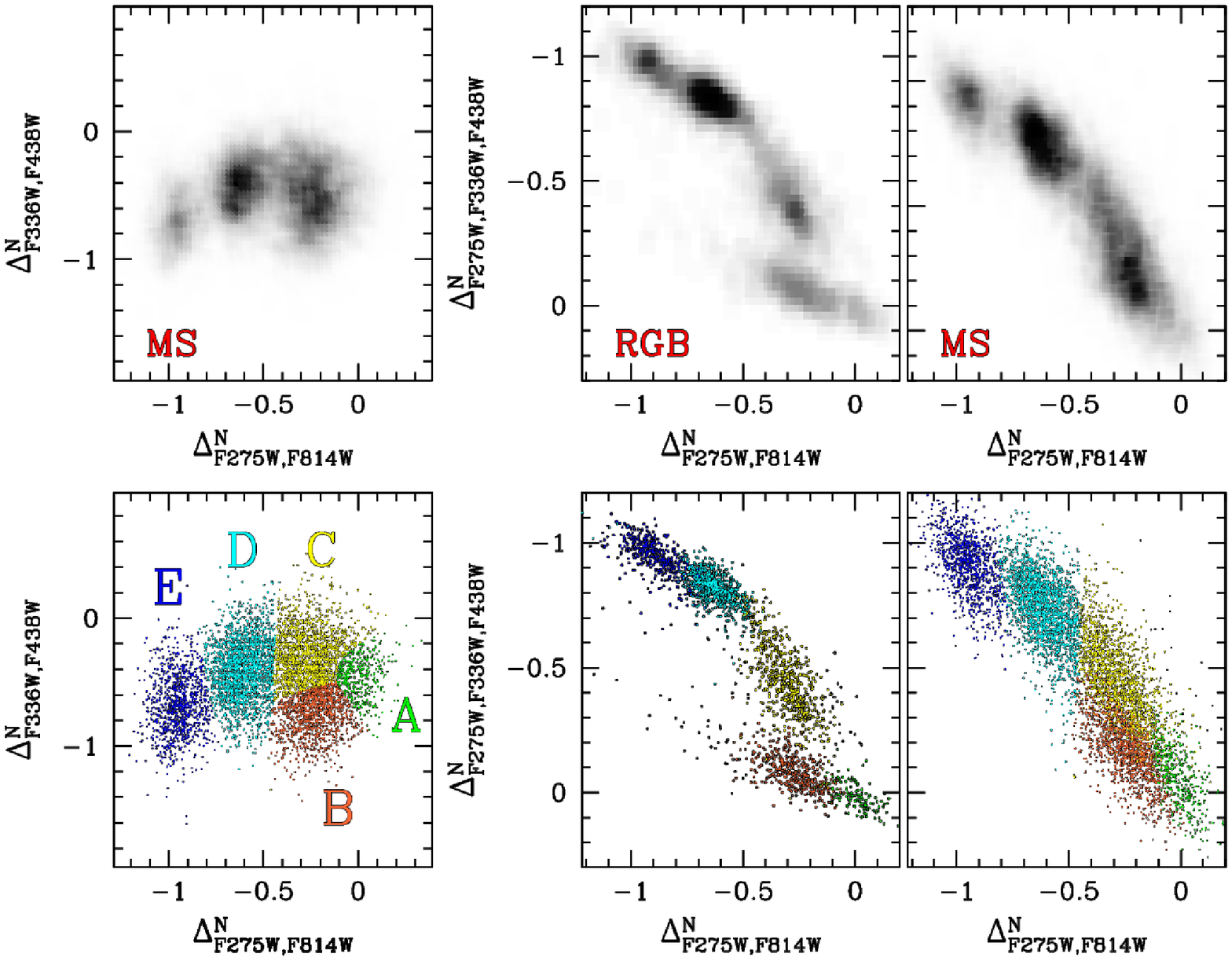}
      \caption{\textit{Left:} Reproduction of the $\Delta_{\rm F336W, F438W}$ vs.\,$\Delta_{\rm F275W, F814W}$ diagram  of Fig.~\ref{NGC2808mMSs2}. Stars in the A, B, C, D, and E groups, defined in this figure, are colored green, orange, yellow, cyan, and blue, respectively (lower panel). The corresponding Hess diagram is shown in the upper-left panel.  Central and middle lower panels show $\Delta^{\rm N}_{\rm F275W, F336W, F438W}$ against \,$\Delta^{\rm N}_{\rm F275W, F814W}$  for RGB and MS stars, respectively while the corresponding Hess diagrams are plotted in the upper panels. }
          \label{seleMS}
   \end{figure}

In order to identify stellar populations along the MS, we exploit the $\Delta^{\rm N}_{\rm F336W, F438W}$ vs.\,$\Delta^{\rm N}_{\rm F275W, F814W}$ diagram and the $\Delta^{\rm N}_{\rm F275W, F336W, F438W}$ vs.\,$\Delta^{\rm N}_{\rm F275W, F814W}$ Hess diagram shown in the lower-left and upper-right panels 
of Fig.~\ref{seleMS}, respectively. 
The distribution of MS stars in this plane is similar to what observed for the RGB, as better highlighted by the Hess diagram of in the upper-right panel Fig.~\ref{seleMS}. There are at least four groups of MS stars that we denominate B, C, D, and E, and colored orange, yellow, cyan, and blue, in the bottom-left and rightmost 
panels, in close analogy with what was done for the RGB. Colors introduced in this figure will be used consistently hereafter. 
Noticeably, the separation among the four groups is less clear for the MS than in the case of the RGB. This could be due to fact that colors of these relatively hot MS stars are less sensitive to light-element variations than the RGB. 
Indeed we have shown in our previous papers that the color difference between multiple MSs and RGBs is due, apart from helium, to different strengths of the molecular bands between  the distinct populations of stars (Marino et al.\,2008; Milone et al.\,2012). In particular, the OH band  and the CH G-band, which are stronger  in the stellar population with the same chemical composition as halo field stars of the same metallicity, mainly fall in the F275W  and the F438W band,  respectively, while the NH band, which is  weaker in  stars of this population, mainly affects the F336W magnitude. 
Population A is not clearly distinguishable, even if a stellar overdensity can be recognized in the Hess diagram at ($\Delta^{\rm N}_{\rm F275W, F814W}$; $\Delta^{\rm N}_{\rm F336W, F438W}$) $\sim$(0.0;$-$0.5). We tentatively associate these stars with population A and color them green in the lower-left panel diagram. 
In the middle- and right-upper panels of Fig.~\ref{seleMS} we compare the  $\Delta^{\rm N}_{\rm F275W, F336W, F438W}$ vs.\,$\Delta^{\rm N}_{\rm F275W, F814W}$ Hess diagrams for RGB and MS stars, while, in the corresponding lower panels, we show the position of Populations A--E in this plane. 

\begin{figure}[htp!]
   \centering
   \epsscale{.75}
      \plotone{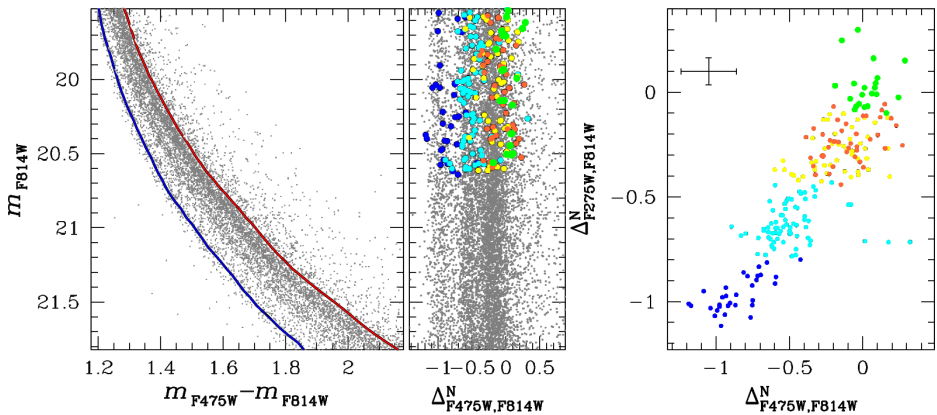}
      \caption{\textit{Left panel:} $m_{\rm F814W}$ against  $m_{\rm F475W}-m_{\rm F814W}$ from the ACS/WFC photometry published by Milone et al.\,(2012a). Red and blue lines are the fiducials of the red and blue MS. \textit{Middle panel:} verticalized $m_{\rm F814W}$ vs.\,$\Delta^{\rm N}_{\rm F475W, F814W}$ diagram. Stars in common with this paper are represented with colored circles.  \textit{Right panel:} $\Delta^{\rm N}_{\rm F275W, F814W}$ (from this paper) vs.\,$\Delta^{\rm N}_{\rm F475W, F814W}$ (from Milone et al.\,2012a).}
          \label{huntMS}
   \end{figure}

In order to further investigate whether MS-A  stars  either correspond to a distinct stellar population or their position in $\Delta^{\rm N}_{\rm F275W, F814W}$ vs.\,$\Delta^{\rm N}_{\rm F336W, F438W}$ plane is entirely due to measurement errors, we adopt a procedure introduced by Anderson et al.\,(2009) and illustrated in Fig.~\ref{huntMS}.
In the left panel we show the $m_{\rm F814W}$ against $m_{\rm F475W}-m_{\rm F814W}$ CMD from Milone et al.\,(2012a). The red and blue lines superimposed on the CMD are the fiducials of the red and the blue MSs and are drawn by hand. 
 The verticalized $m_{\rm F814W}$ vs.\,$\Delta^{\rm N}_{\rm F475W, F814W}$ diagram is plotted in the central panel, while the right panel shows $\Delta^{\rm N}_{\rm F275W, F814W}$  vs.\,$\Delta^{\rm N}_{\rm F475W, F814W}$. Stars in common with this paper are marked with colored circles. 

Photometry by Milone et al.\,(2012a) comes from ACS/WFC images, hence represents a different dataset than the WFC3 ones used in this paper. If the large $\Delta^{\rm N}_{\rm F275W, F814W}$ value for stars in the group A derived  from UVIS/WFC3 photometry is entirely due to photometric errors, then these stars have the same probability of having either small or large $\Delta^{\rm N}_{\rm F475W, F814W}$ derived from WFC/ACS. The systematically large $\Delta^{\rm N}_{\rm F475W, F814W}$ value of stars in the group A, shown in the right panel of Fig.~\ref{huntMS}, is evidence that they belong to a distinct stellar population.

\subsection{Chemical composition of stellar populations}
\label{sub:chem}
\begin{figure}[htp!]
   \centering
   \epsscale{.99}
      \plotone{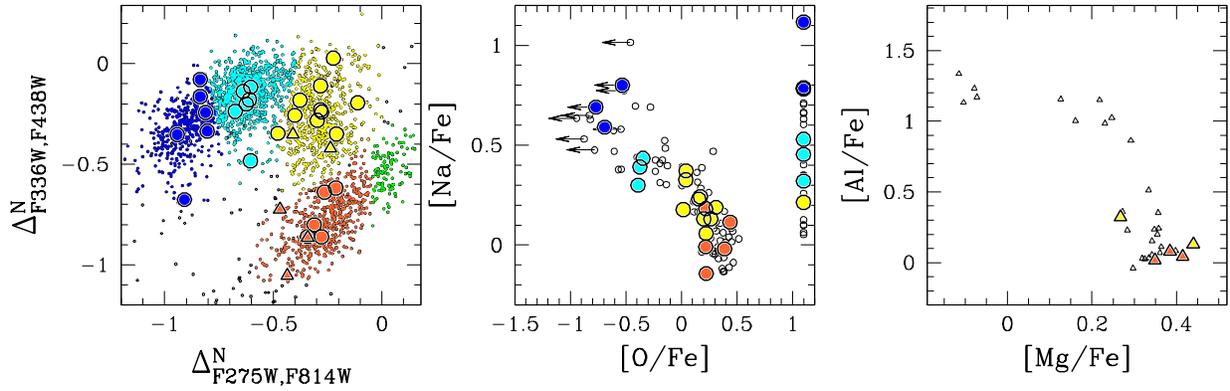}
      \caption{\textit{Left panel:} reproduction of the $\Delta_{\rm F336W, F438W}^{\rm N}$ vs.\,$\Delta_{\rm F275W, F814W}^{\rm N}$ diagram of Fig.~\ref{seleRGBs}. Stars for which spectroscopic measurement are available are marked with large symbols. 
\textit{Middle panel:} sodium-oxygen anticorrelation for RGB stars of NGC\,2808 from Carretta et al.\,(2006). 
Large orange, yellow, cyan, and blue dots indicate spectroscopic targets of population B, C, D, and E, respectively. 
Stars for which only [Na/Fe] measurements are available have been arbitrarily plotted at [O/Fe]=1.1. No population-A stars are present in the Carretta et al.\,sample. \textit{Right panel:} magnesium-aluminum anticorrelation from Carretta\,(2014). Population-B and population-C stars are indicated with orange and yellow triangles, respectively.
}
          \label{NaO}
   \end{figure}

Spectroscopy of RGB stars has revealed that NGC\,2808 exhibits a very extended sodium-oxygen anticorrelation, with [O/Fe] spanning a range of more than 1 dex (Carretta et al.\,2006).
Twenty-seven stars analyzed by Carretta and collaborators are also included in our photometric sample thus providing useful information on the chemical composition of the stellar populations we have identified in the previous sections.
Carretta's stars are marked with large symbols in Fig.~\ref{NaO} where we reproduce the $\Delta_{\rm F336W, F438W}^{\rm N}$ vs.\,$\Delta_{\rm F275W, F814W}^{\rm N}$
diagram of Fig.~\ref{seleRGBs} (left panel), the Na-O anticorrelation from Carretta et al.\,(2006, middle panel), and the Mg-Al anticorrelation from Carretta\,(2014, right panel).
Noticeably, stars in the B, C, D, and E stellar groups defined in this paper have almost the same iron content within $\sim$0.05 dex but populate different regions of the Na-O plane. 
The average elemental abundance for stars of population B, C, D, and E is listed in Table~\ref{tab:abb}, where we also provide the dispersion, $\sigma$, and the number of stars, N, in each population with available abundances. The error is estimated as the $\sigma$ divided by the square root of N$-$1. 
Unfortunately, there are no population-A stars in the sample analyzed by Carretta and collaborators.

The five population-B stars in common with the Carretta et al.\,sample have all primordial Na and O ([Na/Fe]$\sim$0.0, [O/Fe]$\sim$0.3), and the ten population-C stars are slightly enhanced in sodium ([Na/Fe]$\sim$0.2 dex) and depleted in O  ([O/Fe]$\sim$0.15 dex) with respect to population B. Population-E stars all have very low oxygen abundance and high [Na/Fe], while population-D stars are consistent with an intermediate chemical composition. 

More recently, Carretta\,(2014) has  determined  Mg and Al abundances for 31 RGB stars in NGC\,2808 from UVES spectra and detected a very extended Mg-Al anticorrelation with three distinct groups of stars with different content of magnesium and aluminum. 
There are 5 stars in common with Carretta (2014), all Mg-rich and  Al-poor and with similar [Mg/Fe] ratio. On average, the two population-C stars are slightly more Al-rich than the three population-B stars by $\sim$0.2 dex, but a larger sample is needed to establish the significance of  this difference. 
 
 Carretta et al.\,(2010) have defined three groups of stars in NGC\,2808 on the basis of their Na and O abundance. A `primordial' component containing all stars with [Na/Fe]$<\sim$0.19, an `extreme' one with [O/Na]$<$0.9, and an `intermediate' component with intermediate values of Na and O. Our findings show that the component that they designated `primordial' contains at least two different stellar populations, i.e.\,B and C, thus  suggesting that NGC\,2808 has experienced a very complex star-formation history.

\begin{table}[!htp]
\center
\scriptsize {
\begin{tabular}{cccc}
\hline
\hline
 Pop.\, &  abundance & $\sigma$ & \#   \\
\hline
 & [O/Fe] & & \\
\hline  
 A &         ---         &      ---   &      0 \\
 B &     0.30$\pm$0.05  &      0.11  &      5 \\
 C &     0.16$\pm$0.04  &      0.11  &      9 \\
 D &  $-$0.37$\pm$0.02  &      0.02  &      3 \\
 E &  $-$0.66$\pm$0.09  &      0.12  &      3 \\
\hline
 & [Mg/Fe] & & \\
\hline  
 A &         ---         &      ---   &      0 \\
 B &     0.38$\pm$0.02  &      0.03  &      3 \\
 C &     0.35$\pm$0.12  &      0.12  &      2 \\
 D &         ---         &      ---   &      0 \\
 E &         ---         &      ---   &      0 \\
\hline
 & [Al/Fe] & & \\
\hline  
 A &         ---         &      ---   &      0 \\
 B &     0.05$\pm$0.02  &      0.03  &      3 \\
 C &     0.23$\pm$0.13  &      0.13  &      2 \\
 D &         ---         &      ---   &      0 \\
 E &         ---         &      ---   &      0 \\
\hline
 & [Na/Fe]  & &\\               
\hline  
 A &         ---         &      ---   &      0 \\
 B &      0.03$\pm$0.06  &      0.13  &      5 \\
 C &      0.21$\pm$0.03  &      0.09  &     10 \\
 D &      0.40$\pm$0.04  &      0.09  &      6 \\
 E &      0.79$\pm$0.08  &      0.18  &      6 \\
\hline  
 & [Fe/H]  & &\\
\hline  
 A &         ---         &      ---   &      0 \\
 B &   $-$1.13$\pm$0.02  &      0.04  &      5 \\
 C &   $-$1.08$\pm$0.02  &      0.06  &     10 \\
 D &   $-$1.12$\pm$0.02  &      0.04  &      6 \\
 E &   $-$1.10$\pm$0.03  &      0.08  &      6 \\
\hline  
\hline
\end{tabular}
}
\caption{ Average abundance of stars in the five stellar populations of NGC\,2808 defined in this paper. Results are inferred by matching photometry with high-resolution-spectroscopy measurements by Carretta et al.\,(2006) and Carretta\,(2014).}
\label{tab:abb}
\end{table}

\section{Carbon, Nitrogen, Oxygen, and Helium of stellar populations from multiple MSs and RGBs.}
\label{helium}
In order to infer the helium abundance of each stellar population of NGC\,2808 we adapted the method introduced by Milone et al.\,(2012b) in their study of 47\,Tucanae to the case of NGC\,2808. This method is based on the comparison between the observed colors of the multiple sequences and the predictions from appropriate isochrones and the relative synthetic spectra. In this section we will exploit the same technique to estimate the abundance of C, N, and O for the five populations of NGC\,2808.

In Sects.~\ref{sec:RGBs} and~\ref{sub:MS} we have used two-color diagrams from appropriate combinations of  F275W, F336W, F438W, and F814W magnitudes to identify five stellar populations along the RGB and the MS. For simplicity, in the following we will indicate as MS-A---E and RGB-A---E the groups of MS and RGB stars of populations A---E.
 To analyze the behavior of multiple sequences, we have plotted $m_{\rm F814W}$ against $m_{\rm X}-m_{\rm F814W}$, where X=F275W, F336W, F438W, F475W\footnote{Due to the small number of RGB stars for which F475W is available we have not used this filter for the study of multiple RGBs}, and F606W, and determined the MS and the RGB fiducial lines of each population.

\begin{figure}[htp!]
   \centering
   \epsscale{.75}
      \plotone{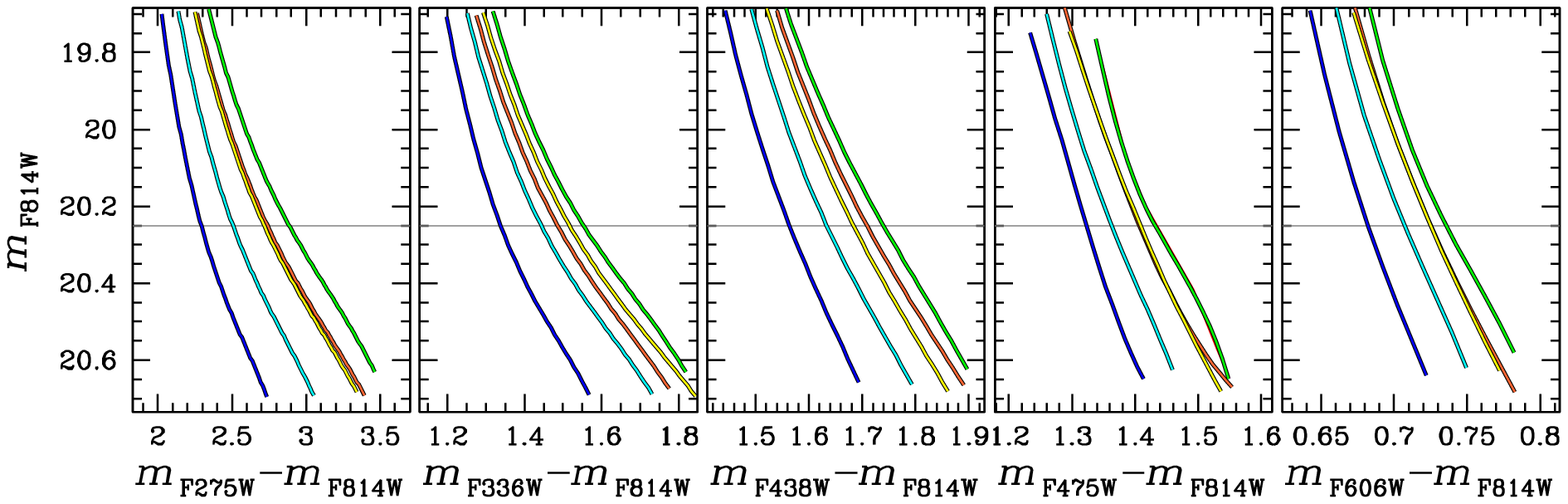}
   \epsscale{.65}
      \plotone{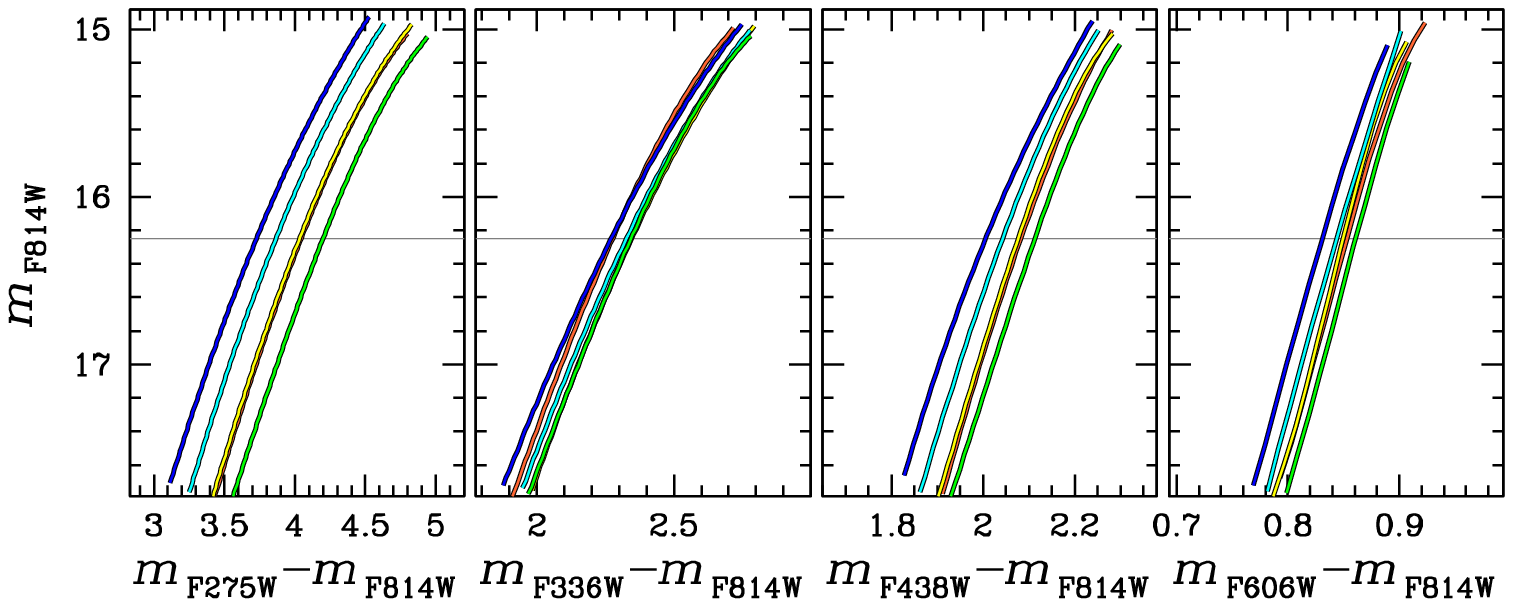}
      \caption{The green, orange, yellow, cyan, and blue lines are the  MS (upper panels) and RGB  (lower panels) fiducial lines for the population A, B, C, D, and E, respectively in the $m_{\rm F814W}$ vs.\,$m_{\rm X}-m_{\rm F814W}$ plane (X=F275W, F336W, F438W, F475W, and F606W). Horizontal gray lines mark the magnitudes at which we have calculated the color distance among the MSs and the RGBs. We have not used the F475W filter for the RGB due to the small number of stars for which we could measure F475W magnitudes.}
          \label{DisMS}
   \end{figure}

Results are shown in Fig.~\ref{DisMS} where we have plotted the RGB and the MS fiducials in the upper and lower panels, respectively.
MS-E exhibits the bluest $m_{\rm X}-m_{\rm F814W}$ color in all the CMDs of Fig.~\ref{DisMS}. Population D has the second bluest MS, while MS-A is redder than the other MSs in all plots. 
MS-B and MS-C are placed between MS-A and MS-D and their relative position changes from one CMD to another. MS-C is slightly bluer than MS-B in $m_{\rm F275W}-m_{\rm F814W}$ and $m_{\rm F438W}-m_{\rm F814W}$ colors but is redder than MS-B in $m_{\rm F336W}-m_{\rm F814W}$. MS-B is almost superimposed on the MS-C in the $m_{\rm F475W}-m_{\rm F814W}$ and $m_{\rm F606W}-m_{\rm F814W}$ colors.  
Lower panels of Fig.~\ref{DisMS} show that the $m_{\rm F275W}-m_{\rm F814W}$, $m_{\rm F438W}-m_{\rm F814W}$, and $m_{\rm F606W}-m_{\rm F814W}$ color order of multiple RGBs and MSs are similar. In the $m_{\rm F814W}$ against $m_{\rm F336W}-m_{\rm F814W}$ CMD, RGB-E and RGB-B share the bluest colors, while RGB-A, RGB-C, and RGB-D
define a red sequence.

\begin{figure}[htp!]
   \centering
   \epsscale{.75}
      \plotone{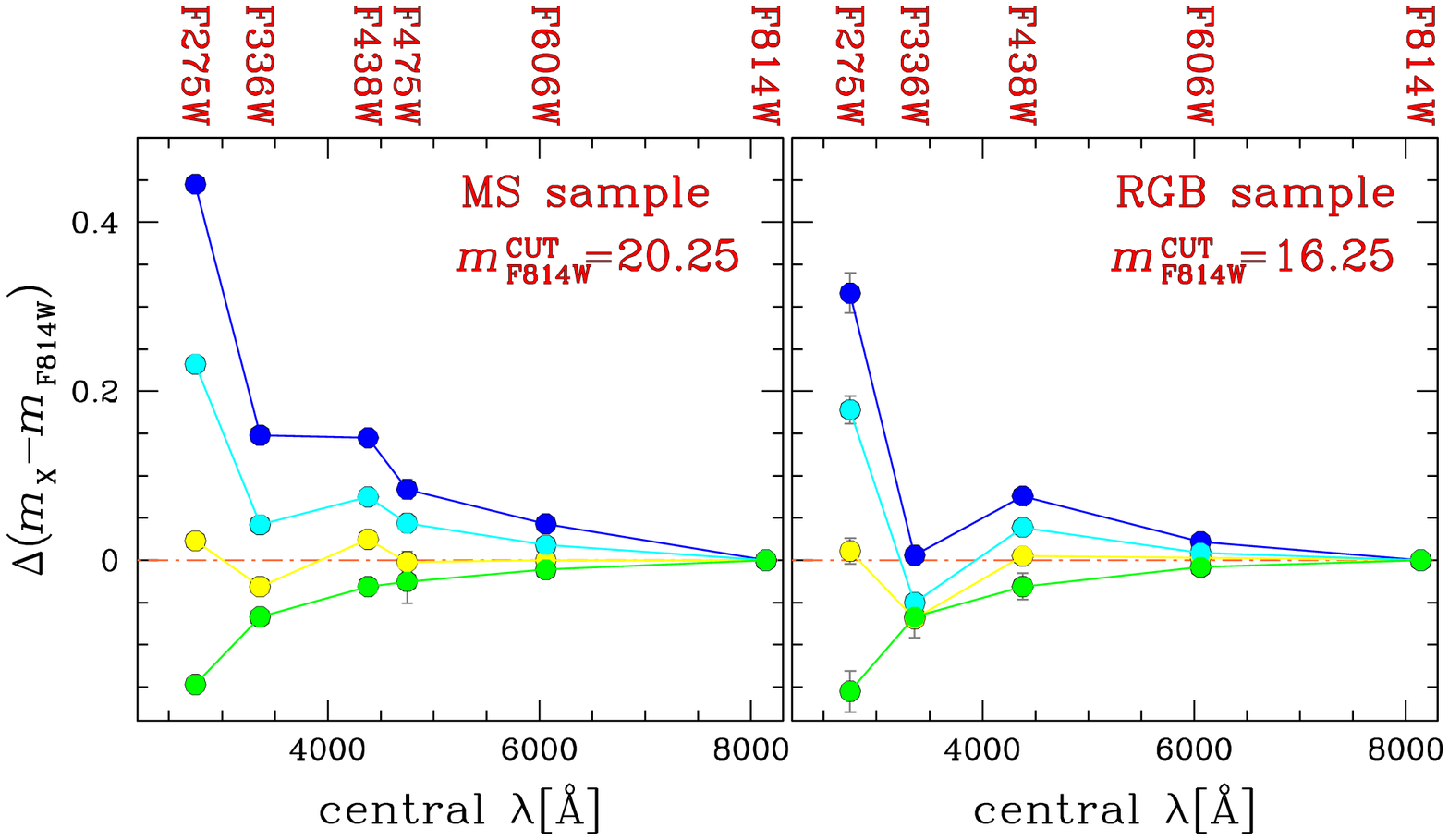}
      \caption{\textit{Left panel:}  $\Delta$($m_{\rm X}-m_{\rm F814W}$)  color distance between MS-B and MS-A, MS-C, MS-D, MS-E (yellow, green, cyan, and blue dots) at $m_{\rm F814W}^{\rm CUT}=$20.25 as a function of the central wavelength of filter $X$.  \textit{Right panel:} Color distance between RGB-B and RGB-A, RGB-C, RGB-D, and RGB-E measured at $m_{\rm F814W}^{\rm CUT}=$16.25 vs.\,the central $\lambda$ of the $X$ filter.}
          \label{DcMS}
   \end{figure}

In order to further investigate multiple MSs and RGBs, we have calculated the $m_{\rm X}-m_{\rm F814W}$ color difference between each MS (or RGB) fiducial and MS-B (or RGB-B) fiducial at a reference magnitude $m_{\rm F814W}^{\rm CUT}$ that we indicate as $\Delta$($m_{\rm X}-m_{\rm F814W}$).
Figure~\ref{DcMS} shows $\Delta$($m_{\rm X}-m_{\rm F814W}$) as a function of the central wavelength of the X filter for MS (left panel, $m_{\rm F814W}^{\rm CUT}=20.25$)  and RGB fiducials (right panel, $m_{\rm F814W}^{\rm CUT}=16.25$).  
We repeated this procedure for $m_{\rm F814W}^{\rm CUT}$=19.80, 19.95, 20.10, 20.40, and 20.55 for the MS and for $m_{\rm F814W}^{\rm CUT}=$15.25, 15.75, 16.75, and 
17.25 for the RGB.
We find that the color separation between populations A and B increases with the color baseline. In the case of both populations D and E, $\Delta$($m_{\rm X}-m_{\rm F814W}$) grows monotonically for X=F606W, X=F475W, and X=F438W, then it drops for X=F336W and reaches its maximum for X=F275W. 
 Population C exhibits almost the same color as population B apart from the case of X=F336W, where population C has negative $\Delta$($m_{\rm X}-m_{\rm F814W}$).

It has been shown that variations in the abundance of both helium and some light elements (e.g.\,C, N, O) are responsible for multiple MSs and RGBs in GCs (e.g.\,Marino et al.\,2008; Sbordone et al.\,2011; Milone et al.\,2012b; Dotter et al.\,2015). In a few GCs multiple sequences are also due to iron variations (e.g.\,Pancino et al.\,2000; Piotto et al.\,2005; Marino et al.\,2012; Paper\,II).
 To investigate the effect of C, N, O, iron and helium on the CMD of NGC\,2808 we started to compute synthetic spectra  that we used as reference for population-B stars (from here on `reference spectrum').
We used BaSTI isochrones (Pietrinferni et al.\,2004, 2009) to estimate the effective temperature ($T_{\rm eff}$=5619 K) and the surface gravity ($\log{g}$=4.56) at  $m_{\rm F814W}=m_{\rm F814W}^{\rm CUT}$ for a MS star with helium ($Y=0.278$).
 We assumed for NGC\,2808 a metallicity of [Fe/H]=$-$1.14 (Harris 1996, 2010 edition), an average reddening E($B-V$)=0.19 (Bedin et al.\,2000) and a distance modulus $(m-M)_{\rm V}$=15.59 (Harris 1996, 2010 edition). 
 We computed a synthetic spectrum with [O/Fe]=0.3, as derived for population B in Sect.~\ref{sub:chem} from Carretta et al.\,(2006) measurements, and [C/Fe]=$-$0.3, and [N/Fe]=0.5 as inferred by Bragaglia et al.\,(2010) for one red-MS star analyzed in their paper. We assumed Y=0.278 for population 
B\footnote{This choice for the helium content of population B  is somehow arbitrary, and in practice adopted to avoid a helium content formally smaller than 
the primordial one for population A. We note that, in this section, we measure helium differences among the different populations, not absolute helium values.
A different choice of primordial helium content for the reference population would have a negligible impact on the estimated $\Delta Y$.}.

We have also computed additional synthetic spectra for stars with the same F814W magnitude, but different chemistry.
We denominated these spectra as `comparison spectra'.
We assumed for each of them the same chemical composition as the reference spectrum, but different abundances for He, C, N, O, and Fe.

Since stars with the same  luminosity in the F814W band, but different He or [Fe/H], have also different temperature and gravity  (e.g., Sbordone et al.\,2011; Cassisi et al.\,2013), when the content of helium and iron are varied, atmospheric parameters are changed accordingly, as predicted by BaSTI isochrones.  
 Specifically, in order to determine the appropriate value of $T_{\rm eff}=T_{\rm eff}^{*}$ for a star with magnitude $m_{\rm F814W}=m^{*}_{\rm F814W}$ and a given content of helium $Y=Y^{*}$, we have calculated the values of $T_{\rm eff, isochrone}$ at $m_{\rm F814W}=m^{*}_{\rm F814W}$ from each available isochrone and estimated $T_{\rm eff}^{*}$ by linearly interpolating the value of $T_{\rm eff, isochrone}$ that corresponds to $Y=Y^{*}$. Interpolations are done within the various available isochrones with helium $Y_{\rm isochrone}$=0.246, 0.28, 0.30, 0.33, 0.35, 0.40.

\begin{figure*}[htp!]
   \centering
   \epsscale{.8}
      \plotone{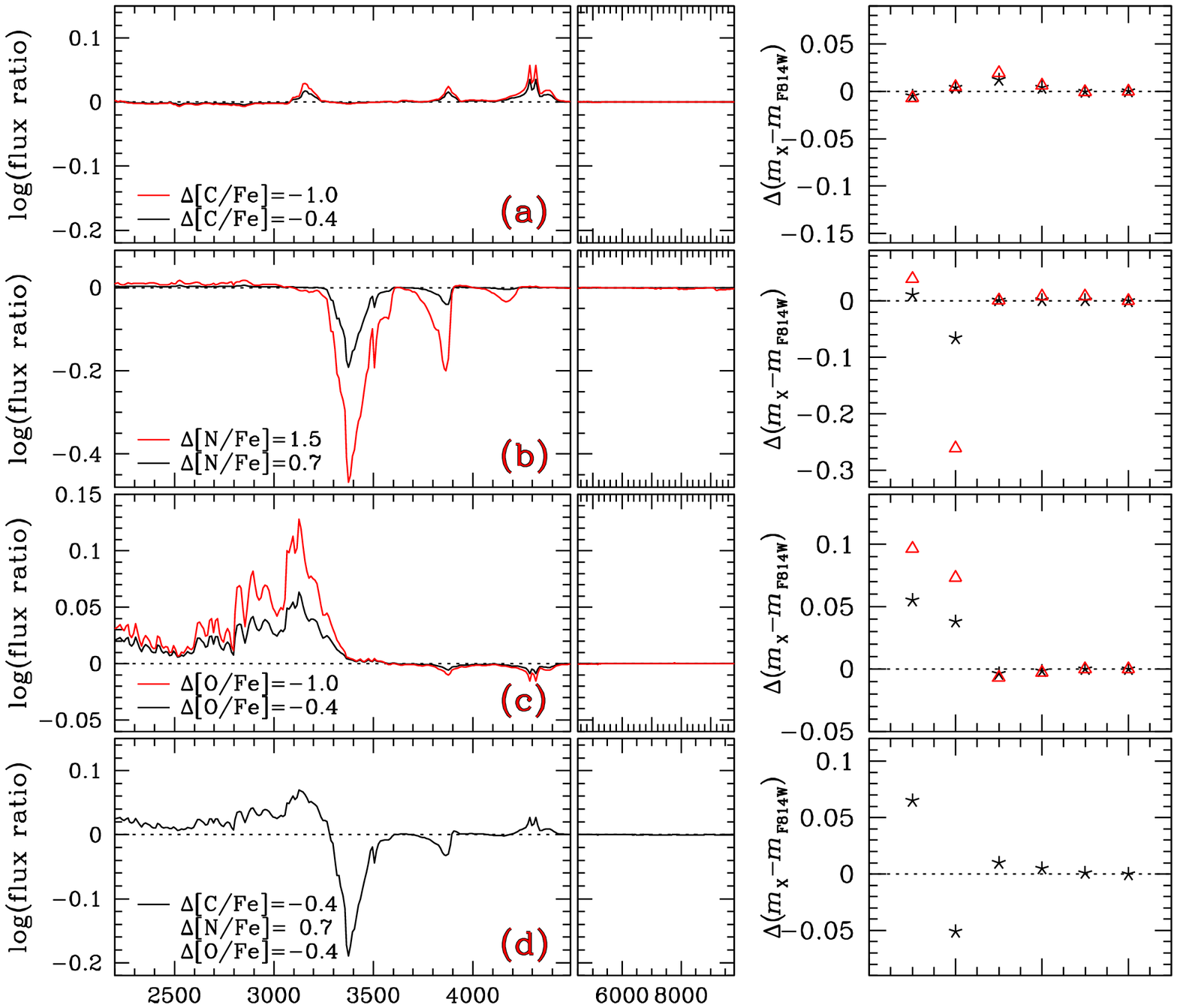}
      \plotone{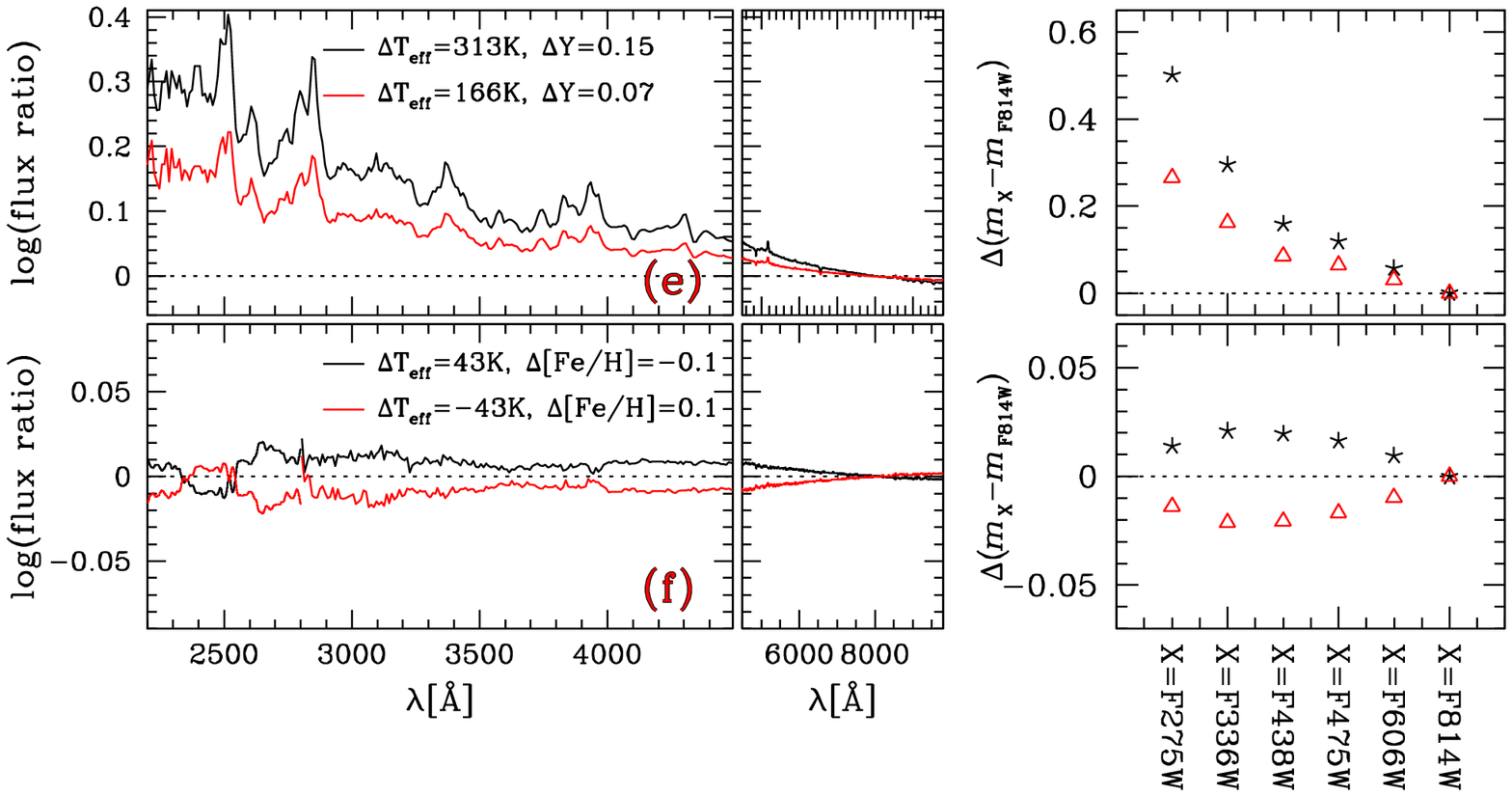}
      \caption{\textit{Left panels:} the ratio between the flux of some comparison spectra and the reference spectrum for a MS-B star at $m_{\rm F814W}^{\rm CUT}=$20.25 is plotted as a function of the wavelength. Each comparison spectrum has the same chemical composition as the reference spectrum apart from the abundance of one element, as indicated in each panel. When $Y$ and [Fe/H] are changed also the atmospheric parameters of the comparison spectrum are varied accordingly, as indicated. \textit{Right panels:} color difference ($\Delta$($m_{\rm X}-m_{\rm F814W}$)) between the comparison spectra and the reference spectrum (see text for details).}
          \label{NGC2808synthtest}
   \end{figure*}

 The adopted range of light elements and iron matches observations from high-resolution spectroscopy (Carretta et al.\,2006; Bragaglia et al.\,2010; Gratton et al.\,2011; Marino et al.\,2014). The difference in temperature, and chemical composition between the comparison and the reference spectrum are indicated in the left panels of Fig.~\ref{NGC2808synthtest}.   

We used ATLAS12 (Kurucz\,2005, Castelli\,2005, Sbordone et al.\,2007) and SYNTHE codes (Kurucz\,2005) to account for the adopted chemical composition and to perform the spectral synthesis in the wavelength interval between 2,000 and 10,000 \AA. Synthetic spectra have been integrated over the transmission curves of the F275W, F336W, F438W filters of UVIS/WFC3 and the F475W, F606W, F814W filters of ACS/WFC, and, for each spectrum, we calculated the color $m_{\rm X}-m_{\rm F814W}$.  

Results are illustrated in the right panels of Fig.~\ref{NGC2808synthtest} and provide an indication on the effect of varying C, N, O, Fe, and He on the observed colors. Left panels show the flux ratio between the comparison spectrum and the reference spectrum  of MS stars with $m_{\rm F814W}^{\rm CUT}=$20.25. The corresponding $\Delta$($m_{\rm X}-m_{\rm F814W}$) color differences are plotted in the right panels.

In the top-left left panels of Fig.~\ref{NGC2808synthtest} we analyze two comparison spectra sharing the same atmospheric parameters and chemical composition as the reference spectrum, but depleted in carbon. A difference in [C/Fe] of 0.4, as observed for red- and blue-MS stars by Bragaglia et al.\,(2010), mainly affects $m_{\rm F438W}-m_{\rm F814W}$ and marginally changes the other colors studied in this work. On the other hand, variations in [N/Fe] produce a large $m_{\rm F336W}-m_{\rm F814W}$ difference (panel b). 
Because of the effect of the OH, NH, and CH bands on the different photometric bands discussed in Sect.~\ref{sub:MS}, changing [O/Fe] mainly affects $m_{\rm F275W}-m_{\rm F814W}$ and $m_{\rm F336W}-m_{\rm F814W}$ colors, with negligible changes
 for visual colors (panels c). 

In panels d we changed C, N, and O in a way that the overall C$+$N$+$O abundance in the reference and in the comparison spectra remains unchanged. In this case we have large differences in $m_{\rm F275W}-m_{\rm F814W}$ and $m_{\rm F336W}-m_{\rm F814W}$.

 As shown by Sbordone et al.\,(2011), helium variation marginally affects the atmospheric structure and the resulting flux distribution. However, stars with the same luminosity but different helium have different effective temperature (see Fig.~1 by Sbordone et al.\,2011, as an example). For this reason, 
 helium variation results in a difference of both visual and ultraviolet colors as shown in panels e, with the $m_{\rm F275W}-m_{\rm F814W}$ and $m_{\rm F336W}-m_{\rm F814W}$ color differences being significantly larger than those in $m_{\rm F438W}-m_{\rm F814W}$ and $m_{\rm F475W}-m_{\rm F814W}$. Also variations in [Fe/H] affect all the colors studied in this paper, but, in this case, the differences corresponding to visual and ultraviolet colors are similar (panels f).

We conclude that the F275W and F336W bandpasses are very sensitive to the detailed chemical abundance of the cluster stars, therefore maximizing the separation among the various sub-sequences due to the variations of light-elements and helium. Specifically, the F275W band is mainly affected by oxygen variations via the strength of the OH molecular bands, while F336W is mostly sensitive to nitrogen via the the NH molecular bands. 
 In contrast, the optical  colors ($m_{\rm F438W}-m_{\rm F814W}$, $m_{\rm F475W}-m_{\rm F814W}$, and $m_{\rm F606W}-m_{\rm F814W}$) are less affected by C, N, O variations, but are very sensitive to temperature variations associated with differences in helium. In addition, a difference in [Fe/H] of less than 0.1 dex, as inferred for NGC\,2808 from high-resolution spectroscopy by Carretta et al.\,(2006), corresponds to small color variations.

Having demonstrated that optical and ultraviolet colors are very sensitive to helium and light-element variations, we have estimated the chemical composition of population E. To do this we have adopted the procedure described above to calculate synthetic spectra with different helium abundances, with Y ranging from 0.246 to 0.400 in steps of $\Delta $Y=0.001, [C/Fe], [N/Fe], and [O/Fe] from $-$2.0 to 2.0 dex in steps of $\Delta$[C/Fe]=0.1; $\Delta$[N/Fe]=0.1, and $\Delta$[O/Fe]=0.1.  The values of $T_{\rm eff}$ and log{g} corresponding to different Y have been derived from isochrones as described above. We used chi-square minimization to determine the best fit between the synthetic colors and observations. The helium, carbon, nitrogen, oxygen difference corresponding to the best-fit model are listed in Table~\ref{table:bf} for different values of $m_{\rm F814W}^{\rm CUT}$ together with its effective temperature and gravity.

 Multi-wavelength photometry of RGB stars also is very sensitive to C, N, O and helium. To better constrain the chemical composition of the five stellar populations of NGC\,2808, we have extended the method above described
for MS stars to the RGB.
 We note that the values of [C/Fe] and [N/Fe] adopted for population-B stars come from spectroscopy of MS stars.  When a star ascend the RGB it is affected by mixing phenomena which alter the original surface abundance of C and N, while keeping constant C$+$N (e.g.\,Iben\,1967). 
To account for this phenomenon, we followed the recipe by Milone\,(2015) and assumed that the reference spectrum for a RGB-B star has a 0.3-dex lower C abundance than that inferred from the MS by Bragaglia et al.\,(2010). Nitrogen abundance has been determined by assuming that the sum C+N remains constant and corresponds to [N/Fe]=0.62 dex. The adopted variation approximately matches the value predicted by Angelou et al.\,(2011) for multiple stellar populations in M\,3.

\begin{figure*}[htp!]
   \centering
   \epsscale{.75}
      \plotone{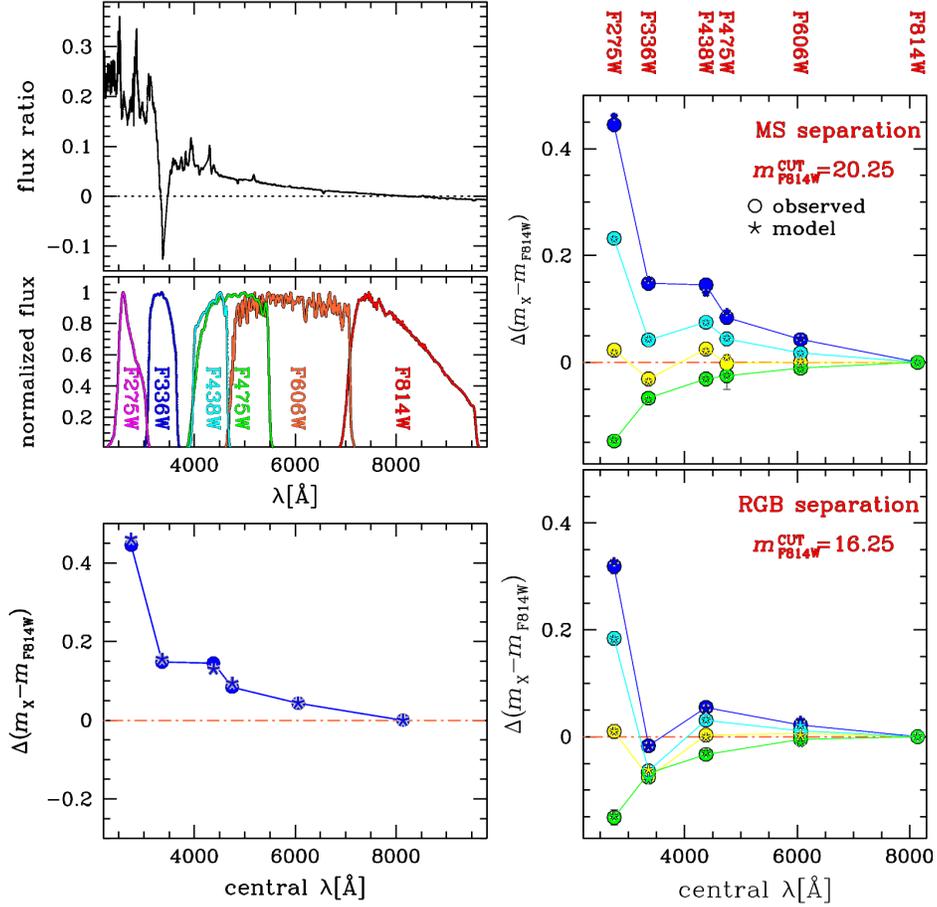}
      \caption{
 \textit{Left Panels:} Flux ratio between the best-fit comparison spectra of population E and population B for MS stars at $m_{\rm F814W}^{\rm CUT}=$20.25 is plotted as a function of the wavelength in the upper panel where we also show the normalized transmission curves of the filters used in this paper.
 Observed  $\Delta$($m_{\rm X}-m_{\rm F814W}$) against the central wavelength of the filter $X$ are plotted with blue circle in the bottom-left panel. Asterisks are inferred from the best-fit synthetic spectra.
 \textit{Right Panels} Reproduction of Fig.~\ref{DcMS} where we plotted $\Delta$($m_{\rm X}-m_{\rm F814W}$) as a function of the central wavelength of filter $X$ as  observed at $m_{\rm F814W}^{\rm CUT}=$20.25 (top) and $m_{\rm F814W}^{\rm CUT}=$16.25 (bottom). In addition, we show the synthetic colors corresponding to the best-fit model (see text for details).}
          \label{NGC2808synthcol}
   \end{figure*}

Results for populations A, C, D, and E are illustrated in Fig.~\ref{NGC2808synthcol}. 
 Upper-left panel shows the flux ratio between the best-fitting MS-E-comparison spectrum and the reference spectrum for $m_{\rm F814W}^{\rm CUT}=$20.25, while the normalized throughput of the filters used in this paper is plotted in the middle-left panel. In the lower-left panel we overplotted on the observed color difference between MS-B and MS-E of Fig.~\ref{DcMS} (blue dots) the color differences derived from synthetic spectra.
 Right panels compare the observed MS and RGB color distance between population B and populations A, C, D, and E of Fig.~\ref{DcMS} and the color differences from the best-fit models of Table~\ref{table:bf} and~\ref{table:bf2}.

From the analysis of multiple MSs, we can  infer that populations D and E are highly helium enhanced by  $\Delta$Y$\sim$0.11 and $\Delta$Y$\sim$0.06, respectively, and are both strongly enhanced in N and depleted in O.  Population C shares almost the same helium as population B ($\Delta$Y$\sim$0.01) and is more nitrogen rich and more oxygen poor than the latter. 
Under the assumption of populations A and B having the same metallicity, population A would have a lower helium content than population B, by $\Delta Y\sim 0.03$, while it would have almost the same nitrogen.

 As suggested by the referee, we have also repeated the helium abundance estimates using only the $m_{\rm F606W}-m_{\rm F814W}$ color difference between each sequence and isochrones with different helium abundance, as this color  is not significantly affected by light-element variations (Sbordone et al.\,2011).  The results are fully consistent with what found from the analysis above, within helium differences less than 0.005.

\begin{figure*}[htp!]
   \centering
   \epsscale{.95}
      \plotone{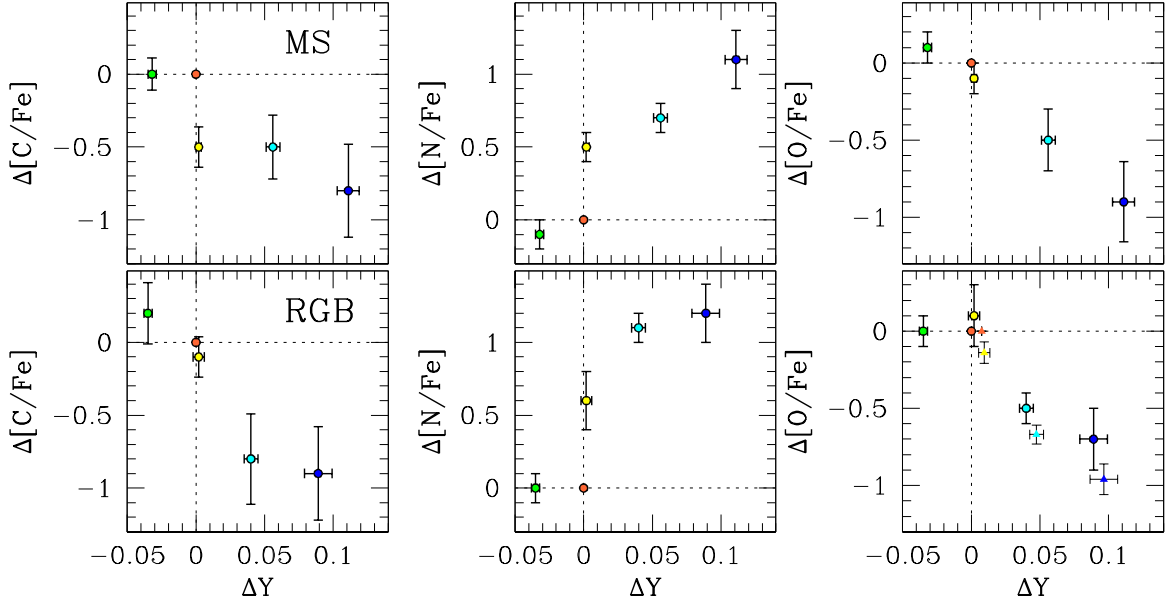}
      \caption{Variation of carbon (left), nitrogen (middle), and oxygen (right) as a function of the helium variation for the five populations of NGC\,2808. Elemental variations are calculated with respect to average abundance of population-B stars. 
 Results obtained from multiple MSs and RGBs are plotted in the upper and lower panels, respectively.
Filled circles indicate the He, C, N, O relative abundances inferred from multi-wavelength photometry, while the values of $\Delta$[O/Fe] inferred from spectroscopy of RGB stars are represented with triangles. For clarity, spectroscopic measurements are shifted by $\Delta$Y$=$0.01.}
          \label{abbFOT}
   \end{figure*}

Results on the estimated compositions of the different populations
are shown in Fig.~\ref{abbFOT}, where we used colored circles to illustrate the 
 differences in carbon (left panel), nitrogen (middle panel), and oxygen (right panel) as a function of the helium variation
 between populations A, C, D, E and population B as inferred from multi-wavelength photometry of multiple MSs (upper panels) and multiple RGBs (lower panels). 
 Overall, we see a correlation between nitrogen and helium, while both [C/Fe] and [O/Fe] anticorrelate with Y. Noticeably there is a large  difference of nitrogen between population B and C associated with a small helium variation.

In the right panel of Fig.~\ref{abbFOT}  colored triangles are obtained from the average oxygen abundance determined in Sect.~\ref{sub:chem} from high-resolution spectroscopy of RGB stars by Carretta et al.\,(2006) for populations B, C, D, and E. In general, results from spectroscopy and photometry are in agreement at one-sigma level, although we note that photometry from this paper systematically
predicts slightly smaller [O/Fe] variations for populations C, D, and E.

\begin{table}
\center
\scriptsize {
\begin{tabular}{lcccccc}
\hline
\hline
Pop.   & $T_{\rm eff}$ & $\log{g}$ & $\Delta$Y &  $\Delta$[C/Fe] &  $\Delta$[N/Fe] & $\Delta$[O/Fe] \\
\hline
 $m_{\rm F814W}^{\rm CUT}$=&19.80  & & & & & \\
\hline
A     &  5830       &   4.49     & $-$0.032   & $-$0.1 & $-$0.1 &     0.1     \\
B     &  5887       &   4.48     &    0.000   &    0.0 &    0.0 &     0.0     \\
C     &  5887       &   4.48     &    0.000   & $-$0.6 &    0.4 &  $-$0.1     \\
D     &  5978       &   4.46     &    0.048   & $-$0.5 &    0.7 &  $-$0.4     \\       %
E     &  6055       &   4.45     &    0.095   & $-$1.2 &    1.0 &  $-$0.9     \\       %
\hline
 $m_{\rm F814W}^{\rm CUT}$=&19.95  & & & & \\
\hline
A     &  5753       &   4.52    & $-$0.028    & 0.0    & $-$0.1      &    0.1  \\
B     &  5805       &   4.51    &    0.000    & 0.0    &    0.0      &    0.0  \\
C     &  5805       &   4.51    &    0.000    & $-$0.6 &    0.5      & $-$0.1  \\
D     &  5902       &   4.50    &    0.052    & $-$0.4 &    0.6      & $-$0.3  \\
E     &  5993       &   4.48    &    0.101    & $-$0.6 &    0.9      & $-$0.5  \\
\hline
 $m_{\rm F814W}^{\rm CUT}$=&20.10  & & & & \\
\hline
A     &  5658       &   4.54    & $-$0.029    &    0.1 & $-$0.1      &    0.1     \\
B     &  5715       &   4.54    &    0.000    &    0.0 &    0.0      &    0.0     \\
C     &  5715       &   4.54    &    0.000    & $-$0.5 &    0.5      & $-$0.1     \\
D     &  5797       &   4.53    &    0.042    & $-$0.5 &    0.6      & $-$0.4     \\    %
E     &  5922       &   4.52    &    0.106    & $-$0.9 &    1.2      & $-$0.7  \\    %
\hline
 $m_{\rm F814W}^{\rm CUT}$=&20.25  & & & & &\\
\hline
A     &  5549       &   4.57    & $-$0.034    &  0.0   & $-$0.1      &    0.2  \\
B     &  5618       &   4.56    &    0.000    &  0.0   &    0.0      &    0.0  \\ %
C     &  5626       &   4.56    &    0.004    & $-$0.6 &    0.5      & $-$0.0  \\ %
D     &  5726       &   4.56    &    0.056    & $-$0.6 &    0.8      & $-$0.6  \\ %
E     &  5832       &   4.55    &    0.112    & $-$1.3 &    1.2      & $-$1.3  \\ %
\hline
 $m_{\rm F814W}^{\rm CUT}$=&20.40  & & & & \\
\hline
A     &  5437       &   4.59    & $-$0.037    &    0.1 & $-$0.1      &    0.2  \\
B     &  5515       &   4.59    &    0.000    &    0.0 &    0.0      &    0.0  \\
C     &  5526       &   4.59    &    0.005    & $-$0.6 &    0.5      & $-$0.1  \\
D     &  5623       &   4.58    &    0.051    & $-$0.8 &    0.9      & $-$0.7  \\
E     &  5747       &   4.58    &    0.110    & $-$1.3 &    1.2      & $-$1.1   \\
\hline
 $m_{\rm F814W}^{\rm CUT}$=&20.55  & & & & &\\
\hline
A     &  5335       &   4.61    & $-$0.033    &    0.0 & $-$0.1      &    0.2  \\
B     &  5406       &   4.61    &    0.000    &    0.0 &    0.0      &    0.0  \\
C     &  5412       &   4.61    &    0.003    & $-$0.7 &    0.6      & $-$0.1  \\
D     &  5526       &   4.61    &    0.056    & $-$0.8 &    0.8      & $-$0.5  \\
E     &  5655       &   4.61    &    0.116    & $-$1.3 &    1.1      & $-$1.0  \\
\hline
 AVERAGE  & & & & &\\
Pop.   &  & $\Delta$Y &  $\Delta$[C/Fe] &  $\Delta$[N/Fe] & $\Delta$[O/Fe] & $\Delta$[(C+N+O)/Fe]\\
\hline
A     &       & $-$0.032$\pm$0.003    &    0.0$\pm$0.1 & $-$0.1$\pm$0.1  &    0.1$\pm$0.1 & 0.1$\pm$0.2 \\
B     &       &    0.000              &    0.0         &    0.0          &    0.0         & 0.0 \\
C     &       &    0.002$\pm$0.002    & $-$0.6$\pm$0.1 &    0.5$\pm$0.1  & $-$0.1$\pm$0.1 & 0.1$\pm$0.1 \\
D     &       &    0.051$\pm$0.005    & $-$0.6$\pm$0.2 &    0.7$\pm$0.1  & $-$0.5$\pm$0.2 & 0.0$\pm$0.1 \\
E     &       &    0.106$\pm$0.008    & $-$1.1$\pm$0.3 &    1.1$\pm$0.2  & $-$0.9$\pm$0.3 & 0.3$\pm$0.2 \\
\hline
\hline
\end{tabular}
}
\caption{Best-fit atmospheric parameters and relative abundances for synthetic spectra of MS stars at different values of $m_{\rm F814W}^{\rm CUT}$. We assumed for the MS-B: Y=0.278, [C/Fe]=$-$0.3, [N/Fe]=0.5, [O/Fe]=0.3. See text for details.}
\label{table:bf}
\end{table}

\newpage
\begin{table}
\center
\scriptsize {
\begin{tabular}{lcccccc}
\hline
\hline
Pop.   & $T_{\rm eff}$ & $\log{g}$ & $\Delta$Y &  $\Delta$[C/Fe] &  $\Delta$[N/Fe] & $\Delta$[O/Fe]\\
\hline
 $m_{\rm F814W}^{\rm CUT}$=&15.25  & & & & & \\
\hline
A     & 4905 & 2.46 & $-$0.031        &    0.4  &    0.1 &    0.1   \\
B     & 4933 & 2.44 &    0.000        &    0.0  &    0.0 &    0.0   \\
C     & 4939 & 2.44 &    0.007        &    0.0  &    0.7 &    0.1   \\
D     & 4963 & 2.43 &    0.033        & $-$0.9  &    1.1 & $-$0.6   \\       %
E     & 5000 & 2.41 &    0.074        & $-$1.0  &    1.4 & $-$0.8   \\       %
\hline
 $m_{\rm F814W}^{\rm CUT}$=&15.75  & & & & & \\
\hline
A     & 5008    &  2.70  & $-$0.035   &    0.0  &    0.0 &     0.2       \\
B     & 5040    &  2.68  &    0.000   &    0.0  &    0.0 &     0.0       \\
C     & 5045    &  2.68  &    0.005   & $-$0.1  &    0.8 &     0.1       \\
D     & 5076    &  2.66  &    0.039   & $-$0.8  &    1.0 &  $-$0.4       \\       %
E     & 5120    &  2.63  &    0.086   & $-$1.1  &    1.2 &  $-$0.7       \\       %
\hline
 $m_{\rm F814W}^{\rm CUT}$=&16.25  & & & & & \\
\hline
A     &  5094   &  2.93  & $-$0.038   &    0.3  & $-$0.1 &     0.3 \\
B     &  5132   &  2.91  &    0.000   &    0.0  &    0.0 &     0.0 \\
C     &  5130   &  2.91  & $-$0.002   & $-$0.2  &    0.7 &     0.0 \\
D     &  5175   &  2.89  &    0.043   & $-$1.0  &    1.1 &  $-$0.4 \\       %
E     &  5232   &  2.86  &    0.100   & $-$1.0  &    1.3 &  $-$0.6 \\       %
\hline
 $m_{\rm F814W}^{\rm CUT}$=&16.75  & & & & & \\
\hline
A     & 5167    &  3.14  & $-$0.034   &    0.1  & $-$0.1 &     0.3 \\
B     & 5205    &  3.13  &    0.000   &    0.0  &    0.0 &     0.0 \\
C     & 5207    &  3.13  &    0.002   & $-$0.2  &    0.4 &     0.0 \\
D     & 5250    &  3.11  &    0.041   & $-$0.7  &    1.0 &  $-$0.5 \\       %
E     & 5307    &  3.09  &    0.092   & $-$0.7  &    1.1 &  $-$0.7 \\       %
\hline
 $m_{\rm F814W}^{\rm CUT}$=&17.25  & & & & & \\
\hline
A     &  5219   &  3.36  & $-$0.037   &    0.0  & $-$0.1 &     0.2     \\
B     &  5265   &  3.35  &    0.000   &    0.0  &    0.0 &     0.0     \\
C     &  5264   &  3.35  & $-$0.001   & $-$0.2  &    0.6 &  $-$0.1     \\
D     &  5316   &  3.33  &    0.041   & $-$0.4  &    1.1 &  $-$0.7     \\       %
E     &  5377   &  3.31  &    0.090   & $-$0.5  &    1.0 &  $-$0.8     \\       %
\hline
\hline
 AVERAGE  & & & & &\\
Pop.   &  & $\Delta$Y &  $\Delta$[C/Fe] &  $\Delta$[N/Fe] & $\Delta$[O/Fe] & $\Delta$[(C+N+O)/Fe]\\
\hline
A     &  ---          & $-$0.035$\pm$0.003    &    0.2$\pm$0.2 &    0.0$\pm$0.1  &    0.2$\pm$0.1 & 0.0$\pm$0.2   \\
B     &  ---          &    0.000              &    0.0         &    0.0          &    0.0         & 0.0   \\
C     &  ---          &    0.002$\pm$0.004    & $-$0.1$\pm$0.1 &    0.6$\pm$0.2  &    0.0$\pm$0.1 & 0.2$\pm$0.2   \\
D     &  ---          &    0.040$\pm$0.005    & $-$0.8$\pm$0.3 &    1.1$\pm$0.1  & $-$0.5$\pm$0.1 & 0.5$\pm$0.1   \\
E     &  ---          &    0.089$\pm$0.010    & $-$0.9$\pm$0.3 &    1.2$\pm$0.2  & $-$0.7$\pm$0.1 & 0.5$\pm$0.2   \\
\hline
\end{tabular}
}
\caption{Best-fit atmospheric parameters and relative abundances for synthetic spectra of RGB stars at different values of $m_{\rm F814W}^{\rm CUT}$. We assumed for the RGB-B: Y=0.278, [C/Fe]=$-$0.6, [N/Fe]=0.62, [O/Fe]=0.3. See text for details.}
\label{table:bf2}
\end{table}

%
\section{RGB bump of the stellar populations}
\label{sec:bump}
In a simple stellar population the luminosity of the RGB bump is an
indicator of its metallicity, age, helium abundance, and C$+$N$+$O content (e.g.\,Cassisi \& Salaris\,1997; Bono et al.\,2001; Bjork \& Chaboyer\,2006; Pietrinferni et al.\,2009, Di Cecco et al.\,2010; Cassisi et al.\,2011).
 In a CMD with multiple sequences as in a GC, the distinct RGB bumps can provide information on the age and the chemistry of the various subpopulations. In the specific case of NGC\,2808, Nataf et al.\,(2013) found that this GC hosts a broadened RGB bump.  Although their photometry was not corrected for differential reddening, and therefore they did not distinguish the different bumps, 
they suggested that this peculiar shape of the RGB bump is likely due to multiple stellar populations with different helium content. 
In this section we investigate the bump of the RGBs identified in Sect.~\ref{sec:RGBs} and infer helium abundance by comparing their observed luminosity with theoretical predictions.
 
\begin{figure*}[htp!]
   \centering
   \epsscale{.75}
      \plotone{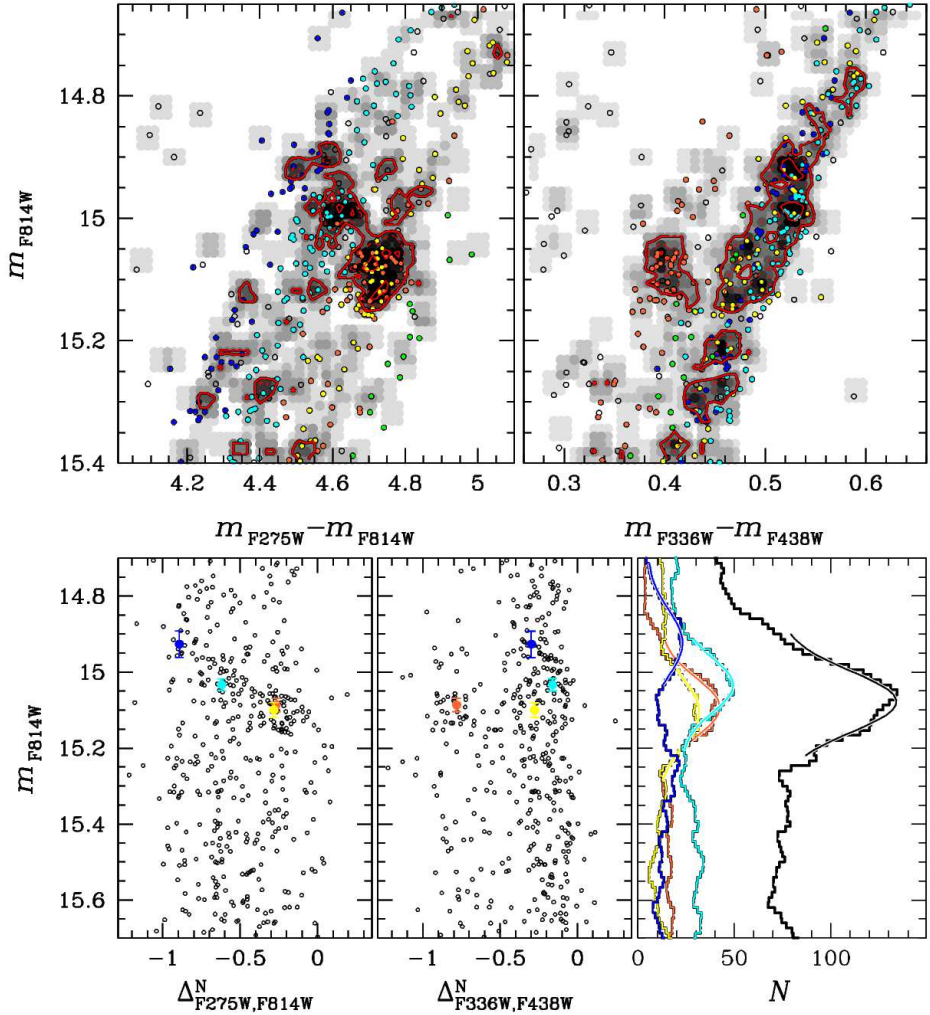}
      \caption{\textit{Upper panels:} $m_{\rm F814W}$  vs.\,$m_{\rm F275W}-m_{\rm F814W}$ (left) and $m_{\rm F814W}$  vs.\,$m_{\rm F336W}-m_{\rm F438W}$ Hess diagrams for stars around the RGB bump. Red iso-density contours are superimposed on each diagram. \textit{Lower panels:} zoom of the $m_{\rm F814W}$ vs.\,$\Delta_{\rm F275W, F814W}^{\rm N}$ (left) and $m_{\rm F814W}$ vs.\,$\Delta_{\rm F336W, F438W}^{\rm N}$ (middle) diagram around the RGB bump. The histograms of the F814W magnitude distribution are plotted  in the right panel for all the stars (black histogram) and for the four  RGBs (colored histograms). Colored circles mark the position of the bump for each population. Continuous lines superimposed on the histograms are the best-fitting Gaussians.}
          \label{NGC2808bumps}
   \end{figure*}

The $m_{\rm F814W}$ against $m_{\rm F275W}-m_{\rm F814W}$ and the $m_{\rm F814W}$ against $m_{\rm F336W}-m_{\rm F438W}$ Hess diagrams of RGB stars plotted in Fig.~\ref{NGC2808bumps} show  that the RGB bump of NGC\,2808 exhibits a complex structure.
 The Hess diagrams reveal multiple bumps, with different F814W luminosities
that are associated with the different stellar populations. 
Lower-left and lower-middle panels of Fig.~\ref{NGC2808bumps} reproduce the same $m_{\rm F814W}$ vs.\,$\Delta_{\rm F275W, F814W}^{\rm N}$ and $m_{\rm F814W}$ vs.\,$\Delta_{\rm F336W, F438W}^{\rm N}$ diagrams of Fig.~\ref{seleRGBs}, but zoomed-in around the RGB bump. In the lower-right panel we plot the histogram of the distribution in $m_{\rm F814W}$ for all the RGB stars shown in the left and in the middle panels (black histogram). Colored histograms represent the $m_{\rm F814W}$ distribution of stars for all the stellar populations of NGC\,2808 we have identified with the exception of population A, which includes too few members for an accurate identification of the bump magnitude. 
Histograms have been obtained by adapting the  naive estimator (Silverman\,1986).  Briefly, we have first defined a regular sample of points ($m_{\rm F814W}^{\rm i}$), ranging from 14.7 to 15.4, with $m_{\rm F814W}^{\rm i+1}-m_{\rm F814W}^{\rm i}=0.2$ mag. Then we have extracted the histogram by associating with each point, $m_{\rm F814W}^{\rm i}$, the number of RGB stars with $m_{\rm F814W}^{\rm i}-\omega/2 \leq m_{\rm F814W}< m_{\rm F814W}^{\rm i}+\omega/2$. We assumed $\omega=$0.013 mag.

In order to obtain the $m_{\rm F814W}$ luminosity of the bumps, we used the following procedure. We started to estimate by eye a raw position for the bump luminosity in each histogram and selected all the points within 0.15 F814W magnitudes from this position. We least square fitted these points with a Gaussian and considered the position of the center of the best-fitting Gaussian ($m_{\rm F814W, bump}$) as a new estimate for the bump luminosity. Finally, we selected all the points of the histogram within 2$\sigma$ (where $\sigma$ is the width of the best-fitting Gaussian) from $m_{\rm F814W, bump}$, for a new least-squares Gaussian fit.
This later determination of $m_{\rm F814W, bump}$ corresponds to our best estimate of the $m_{\rm F814W}$ luminosity of the bump.
Errors on $m_{\rm F814W, bump}$ are determined by bootstrapping with replacements performed 1,000 times on the sample of analyzed RGB stars.
The error bars indicate the 1$\sigma$ (68.27$^{\rm th}$) percentile of the bootstrapped measurements. Colored dots in the left and middle panel of Fig.~\ref{NGC2808bumps} mark the position of the bumps for the four stellar populations in the $m_{\rm F814W}$ vs.\,$\Delta_{\rm F275W, F814W}^{\rm N}$ and $m_{\rm F814W}$ vs.\,$\Delta_{\rm F336W, F438W}^{\rm N}$ diagrams. We repeated the same procedures to estimate the bump position in the F275W, F336W, F438W, and F814W bands.
The RGB-bump magnitudes ($m_{\rm X, bump}^{\rm B,C,D,E}$, X=F275W, F336W, F438W, F606W, and F814W) for four stellar populations are listed in
Table~\ref{tab:bumps}, while in Fig.~\ref{bumps} we show the difference between the RGB-bump magnitude of populations C, D, and E and the RGB-bump magnitude of population B against the central wavelength of the X filter.

In four bands, namely F275W, F438W, F606W, and F814W, the RGB-bump brightness anti-correlates with $\Delta_{\rm F275W, F814W}^{\rm N}$ (e.g.\,lower left panel of Fig.~\ref{NGC2808bumps}). Population E hosts the brightest bump. The RGB bumps of populations B and C share almost the same luminosity, population-D bump is brighter than those of populations B and C. The magnitude separation among the bumps is nearly constant in F438W, F606W, and F814W but increases by a factor of $\sim$2 in F275W, as shown in Fig.~\ref{bumps}. We also note that the order of the RGB-bump brightness is different in F336W, where population E still has extreme values of $m_{\rm F336W, bump}$, but the RGB bumps of both populations C and D are fainter than that of population B.

   \begin{figure}[ht!]
   \centering
   \includegraphics[width=10 cm]{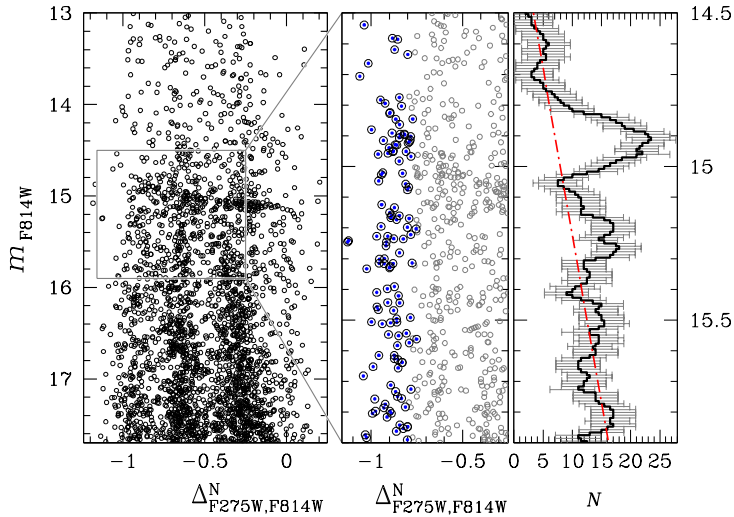}
   \caption{ \textit{Left panel:}  Reproduction of the $m_{\rm F814W}$ vs.\,$\Delta_{\rm F275W, F814W}^{\rm N}$ diagram for RGB stars of Fig.~\ref{NGC2808pops}. \textit{Middle panel:} Zoom of left-panel diagram around the bump of population-E, where RGB-E stars are marked with blue dots. \textit{Right panel:} Histogram of the F814W magnitude distribution. Grey error bars are Poisson errors while the red dashed-dotted line is the assumed luminosity function for the RGB (see text for details).}
         \label{rep1}
   \end{figure}

 To investigate the significance of the RGB bumps of populations B--E we have performed the analysis illustrated in Fig.~\ref{rep1} for population E.
 The left panel of Fig.~\ref{rep1} reproduces the verticalized $m_{\rm F814W}$ vs.\,$\Delta_{\rm F275W, F814W}^{\rm N}$ diagram of RGB stars shown in Fig.~\ref{NGC2808pops}, while the middle panel is a zoom of the region around the bump of population E.
 The histogram in the right panel of Fig.~\ref{rep1} reproduces the F814W magnitude distribution for RGB-E stars (blue dots in the middle panel) already plotted in Fig.~\ref{NGC2808bumps}. The red line is the best fit straight line for the luminosity function of the upper-RGB and has been obtained by excluding stars with $14.7<m_{\rm F814W}<15.1$ to avoid the contamination from the bump.
 Grey error bars are Poisson errors calculated as the square root of the number of stars used to derive each point in the histogram minus one.

Then we have simulated 1,000 verticalized CMDs for RGB stars by assuming same magnitude distribution as predicted by the red line, such that each synthetic diagram has the same number of RGB stars as in the observed CMD.
 We have applied to each CMD the same procedure to derive the F814W magnitude distribution as for real stars. 
 For each simulation we have calculated the difference between the area of the histogram derived from observations and the simulated one.\\
$\Delta Area=\sum_{i}^{14.8<m_{\rm F814W}<15.1}{N_{\rm i}^{\rm observed}-N_{\rm i}^{\rm simulated}}$\\
and find that $\Delta Area$ is significantly larger than zero in all the simulations. We extended the analysis to the bump of populations B, C, and D and find similar results thus concluding that the overdensities of stars resulting from the analysis illustrated in Fig.~\ref{NGC2808bumps} are very likely the RGB bumps of populations B--E. 

In an attempt to interpret observations on the RGB bumps in the four populations we compare them with models.
 Specifically, we have used two different sets of isochrones from BaSTI (Pietrinferni et al.\,2004, 2006, 2009) and from Ventura et al.\,(1998, 2009, hereafter Roma models). 

We have estimated the luminosity difference between the RGB bump of helium-enhanced (Y=0.30, 0.35, 0.37) and helium-normal (Y=0.248) synthetic CMDs by using the same procedure described above for real stars and compared these magnitude differences with observations. 
 Synthetic CMDs have been generated by using ASs and isochrones from both BaSTI and Roma models. We limited our study to the F814W filter which is only marginally affected by variations in light element abundances at fixed Z. 
 We interpolated the value of $Y$ which matches the observed $\Delta m_{\rm X, bump}^{\rm C,D,E}$ and assumed the corresponding value of $Y$ as the best estimate of the helium abundance of populations C, D, and E.  

The $\Delta Y$ estimated from the difference in luminosity of the bump using both BaSTI and Roma models are listed in Table~\ref{tab:bumps}.  We assumed that the stellar populations are coeval, as suggested by Piotto et al.\,(2007) who have revealed a narrow SGB in the analyzed $m_{\rm F814W}$ vs.\,$m_{\rm F475W}-m_{\rm F814W}$ CMD, and used two different  values for the absolute ages: 11.5 Gyr and 10.0 Gyr.
We find that both population D and E are helium-enhanced with respect to population B, while population B and C share almost the same helium content, in analogy with what was inferred from multiple RGBs and MSs in Sects.~\ref{sec:RGBs} and~\ref{sub:MS}. 
%

   \begin{figure}[htp!]
   \centering
   \epsscale{.7}
      \plotone{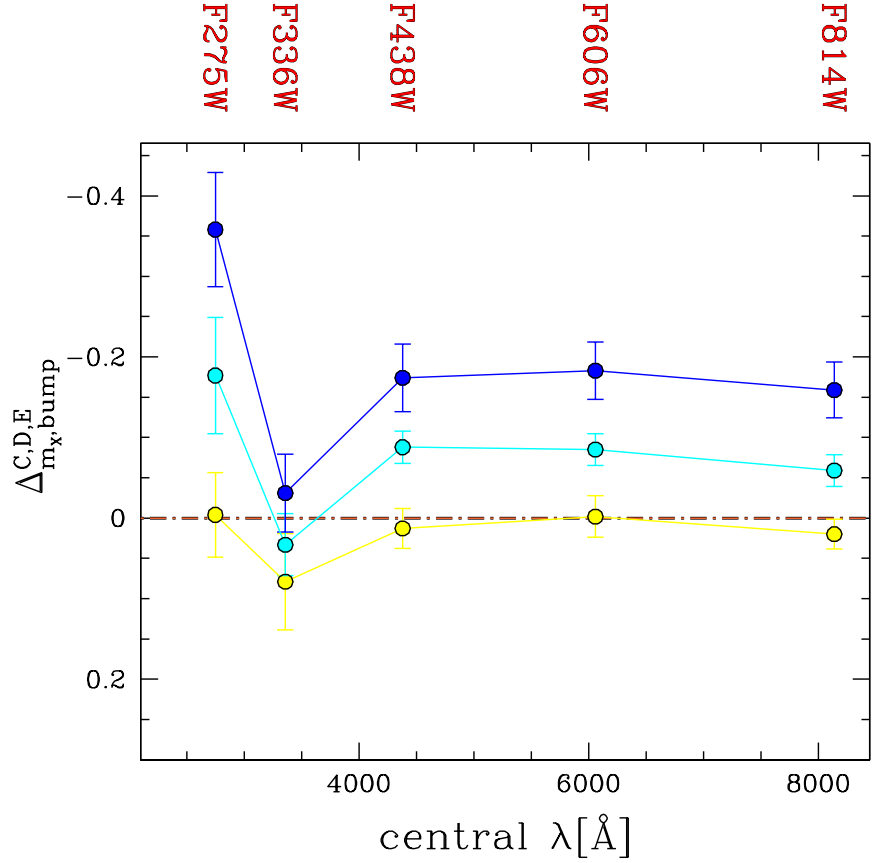}
      \caption{Yellow, cyan, and blue dots indicate the observed magnitude difference between the RGB-bump magnitude of populations C, D, and E ($m_{\rm X, bump}^{\rm C,D,E}$, where X=F275W, F336W, F438W, F606W, and F814W), respectively, and the RGB-bump magnitude of population B ($m_{\rm X, bump}^{\rm B}$).} 
          \label{bumps}
   \end{figure}

\begin{table}[!htp]
\center
\scriptsize {
\begin{tabular}{ccccc}
\hline
\hline
X  & $m_{\rm X, bump}^{\rm B}$ & $m_{\rm X, bump}^{\rm C}$ & $m_{\rm X, bump}^{\rm D}$& $m_{\rm X, bump}^{\rm E}$ \\
\hline
F275W & 19.95$\pm$0.05 & 19.94$\pm$0.02 & 19.77$\pm$0.05 & 19.59$\pm$0.05 \\
F336W & 17.75$\pm$0.03 & 17.83$\pm$0.06 & 17.78$\pm$0.03 & 17.72$\pm$0.04 \\
F438W & 17.35$\pm$0.02 & 17.36$\pm$0.02 & 17.26$\pm$0.02 & 17.18$\pm$0.04 \\
F606W & 15.99$\pm$0.02 & 15.99$\pm$0.02 & 15.91$\pm$0.02 & 15.81$\pm$0.03 \\
F814W & 15.07$\pm$0.02 & 15.09$\pm$0.02 & 15.01$\pm$0.02 & 14.91$\pm$0.03 \\
\hline
 Model  & $\Delta Y_{\rm B}$ & $\Delta Y_{\rm C}$  & $\Delta Y_{\rm D}$ & $\Delta Y_{\rm E}$ \\
\hline
BaSTI, 10.0 Gyr   &   0.000   &  $-$0.010$\pm$0.008  &  0.029$\pm$0.012 &  0.103$\pm$0.025 \\ 
BaSTI, 11.5 Gyr   &   0.000   &  $-$0.008$\pm$0.008  &  0.026$\pm$0.012 &  0.098$\pm$0.025 \\ 
Roma, 10.0 Gyr    &   0.000   &  $-$0.009$\pm$0.013  &  0.035$\pm$0.013 &  0.100$\pm$0.023 \\ 
Roma, 11.5 Gyr    &   0.000   &  $-$0.007$\pm$0.013  &  0.030$\pm$0.013 &  0.085$\pm$0.022 \\ 
\hline
\hline
\end{tabular}
}
\caption{
Observed $m_{\rm X}$ luminosity of RGB bump (X=F275W, F336W, F438W, F606W, F814W), and helium difference with respect to population B inferred from the F814W luminosity of the bump.}
\label{tab:bumps}
\end{table}

\section{Comparison among the $\Delta Y$ from the different methods}\label{comparison}

\begin{figure}[htp!]
   \centering
   \epsscale{.7}
      \plotone{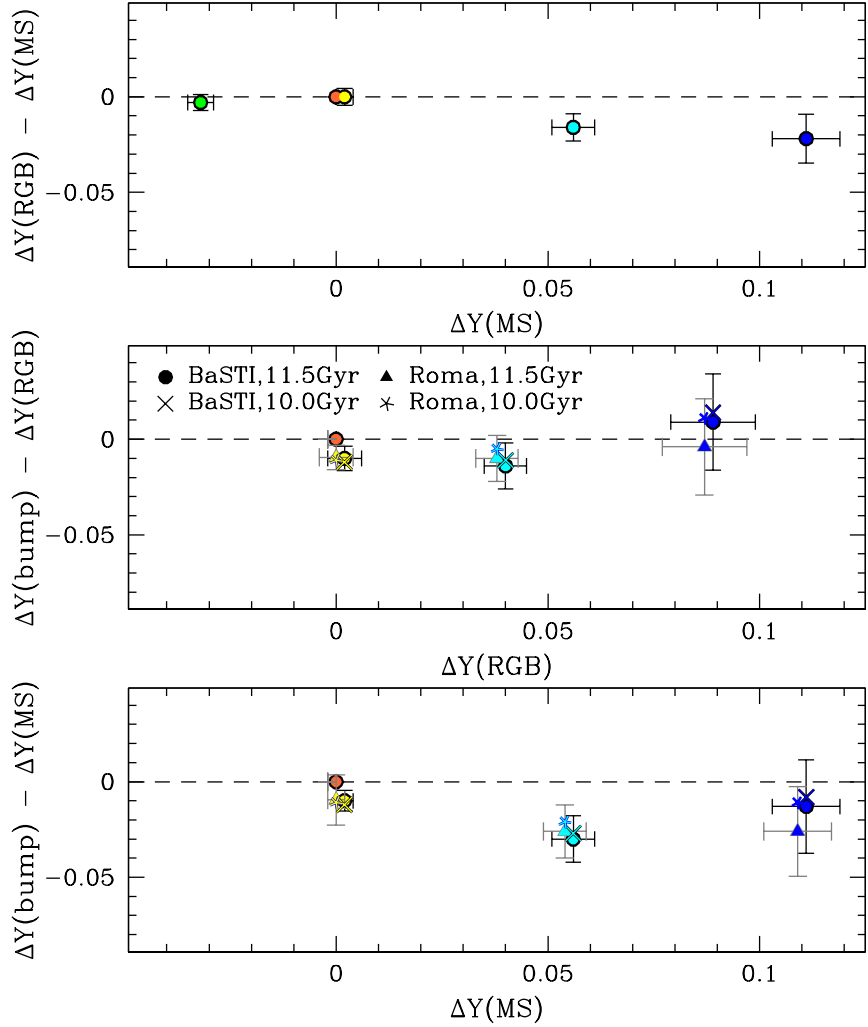}
      \caption{ \textit{Upper panel:} Difference between the $\Delta$Y estimated from the RGB  and the MS for populations A, B, C, D, E. 
                        \textit{Middle panel:} Difference  between the $\Delta$Y coming from the bump location and the RGB.
                        \textit{Lower panel:} Difference between the $\Delta$Y coming from the bump location and the MS. 
                        We used two different isochrone sets and two different ages to estimate $\Delta$Y from the magnitude of the bump. 
                        } 
          \label{comparebump}
   \end{figure}

Figure~\ref{comparebump}  shows the difference between the $\Delta Y$ estimated from the RGB  and the MS (upper panel), the bump location and the RGB (middle panel), and the bump location and the MS (lower panel).  We note that there is a small discrepancy between the $\Delta Y$ from the MS and the RGB for populations D and E, thought within $\leq$2$\sigma$. The $\Delta Y$ from the RGB is in fair agreement with the value calculated using the bump luminosity (within 1 $\sigma$). The comparison of the middle panel of Fig.~\ref{comparebump}  would favour a younger isochrones. 

Note that several authors have found that NGC 2808 has younger (10-15\%) age with respect to the average age of intermediate-metallicity cluster  (Rosenberg et al.\,1999; De Angeli et al.\,2005; VandenBerg et al.\,2013; Milone et al.\,2014). The comparison between the $\Delta Y$ from the MS and the bump is
somehow less satisfactory, though still within 1-2 $\sigma$ (depending on the adopted models). Again, a younger age seems to be favoured.
A discussion on the origin of these discrepancies is beyond the purposes of the present paper and will be postponed to
a future paper. 

\section{The Horizontal Branch}
\label{sec:HB}
Among Galactic GCs, NGC\,2808 hosts one of the most extended HBs, which spans an extreme  $m_{\rm F606W}-m_{\rm F814W}$ color range ($L2$=$\sim$0.9, Milone et al.\,2014). The distribution of stars along the HB of this cluster exhibits three significant gaps, which separate four groups of HB stars. A red HB hosting about half of the total number of HB stars, and three distinct segments of blue HB stars  (Sosin et al.\,1997; Bedin et al.\,2000; Piotto et al.\, 2007). 
The CMD shown in Fig.~\ref{SummaryNGC2808} confirms this complex morphology.

 Spectroscopy of HB stars in NGC\,2808 further reveals that stars with different light-element abundance are distributed along different HB regions. Red-HB stars are Na-poor and O-rich, while blue-HB stars are depleted in oxygen and enhanced in sodium (Gratton et al.\,2011; Marino et al.\,2014).

\begin{figure}[htp!]
   \centering
   \epsscale{.75}
      \plotone{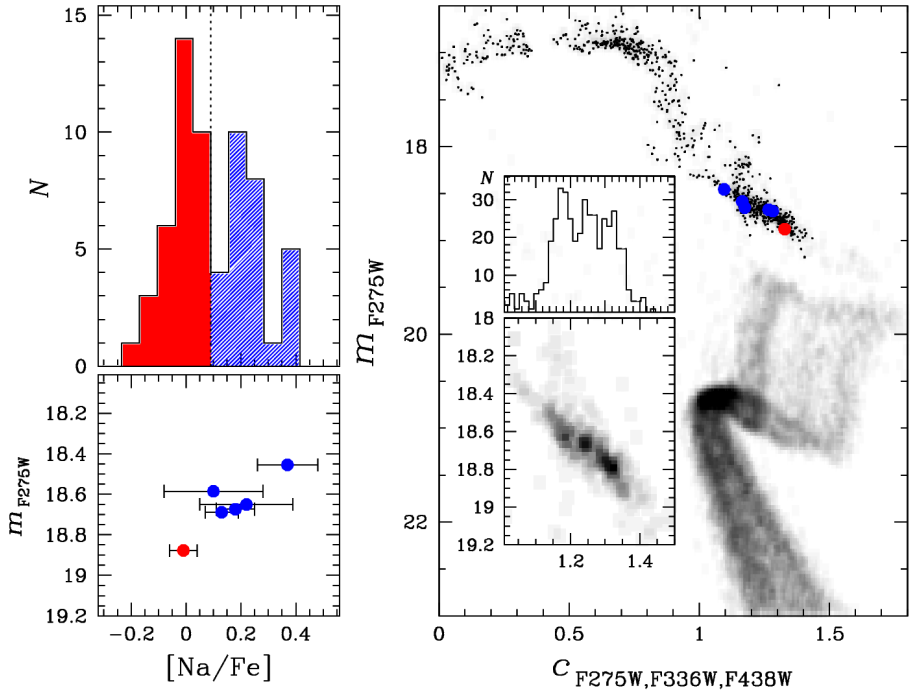}
      \caption{ \textit{Upper-left panel:} histogram of the distribution of [Na/Fe] for red HB stars (Marino et al.\,2014).  \textit{Lower-left panel:} $m_{\rm F275W}$ vs.\,[Na/Fe]. \textit{Right panel:} $m_{\rm F275W}$ vs.\,$C_{\rm F275W, F336W, F438W}$ Hess-diagram. The Hess diagram in the inset inset is a zoom around the red HB and the histogram of the distribution in $C_{\rm F275W, F336W, F438W}$ for red-HB stars is also shown. Na-rich and Na-poor stars defined by Marino and collaborators are colored blue and red, respectively.}
          \label{speHB}
   \end{figure}

 The sodium distribution of red-HB stars is bimodal, with Na-rich stars having, on average, bluer $B-V$ and $U-V$ colors (Marino et al.\,2014).  The histogram of the [Na/Fe] distribution for red HB stars from Marino et al.\,(2014) is reproduced in the upper-left panel of Fig.~\ref{speHB}, where we have colored red and blue the two stellar populations identified by these authors.
 For six HB stars in the {\it HST} field of view sodium abundances are available  from Marino and collaborators. There is a clear anti-correlation between [Na/Fe] and  $m_{\rm F275W}$, with sodium-poor stars having also fainter luminosity in F275W as shown in the lower-left panel of Fig.~\ref{speHB}.

To further investigate the connection between stellar populations with different sodium abundance and the red HB we combine optical and ultraviolet photometry.
Since $m_{\rm F275W}-m_{\rm F336W}$ and $m_{\rm F336W}-m_{\rm F435W}$ colors are very efficient in separating stellar populations along the red HB of GCs, in the right panel of Fig.~\ref{speHB} we plot the $m_{\rm F275W}$ vs.\,$C_{\rm F275W, F336W, F438W}$ Hess diagram for NGC\,2808. The fact that stars with different [Na/Fe] populate different regions of the $m_{\rm F275W}$ vs.\,$C_{\rm F275W, F336W, F438W}$ diagram supports the conclusions by Gratton et al.\,(2011) and Marino et al.\,(2014) that the red HB of NGC\,2808 is not consistent with a simple stellar population.

 The inset shows a zoom around the red HB. The distribution of stars along the HB is multimodal, with three main groups of stars clustered around $C_{\rm F275W, F336W, F438W} \sim$ 1.15, 1.25, and 1.33 as highlighted by the histogram distribution of the pseudo-color $C_{\rm F275W, F336W, F438W}$ shown in the inset. 
An appropriate comparison with HB theoretical models is required to disentangle the effect of mass loss, evolved stars, and multiple stellar populations on the morphology of the red HB and to understand whether the three bumps correspond to distinct populations, but this is beyond the purposes of the present paper.
\section{Multiple populations along the AGB}\label{AGB}
\label{sec:agb}
\begin{figure}[htp!]
   \centering
   \epsscale{.75}
      \plotone{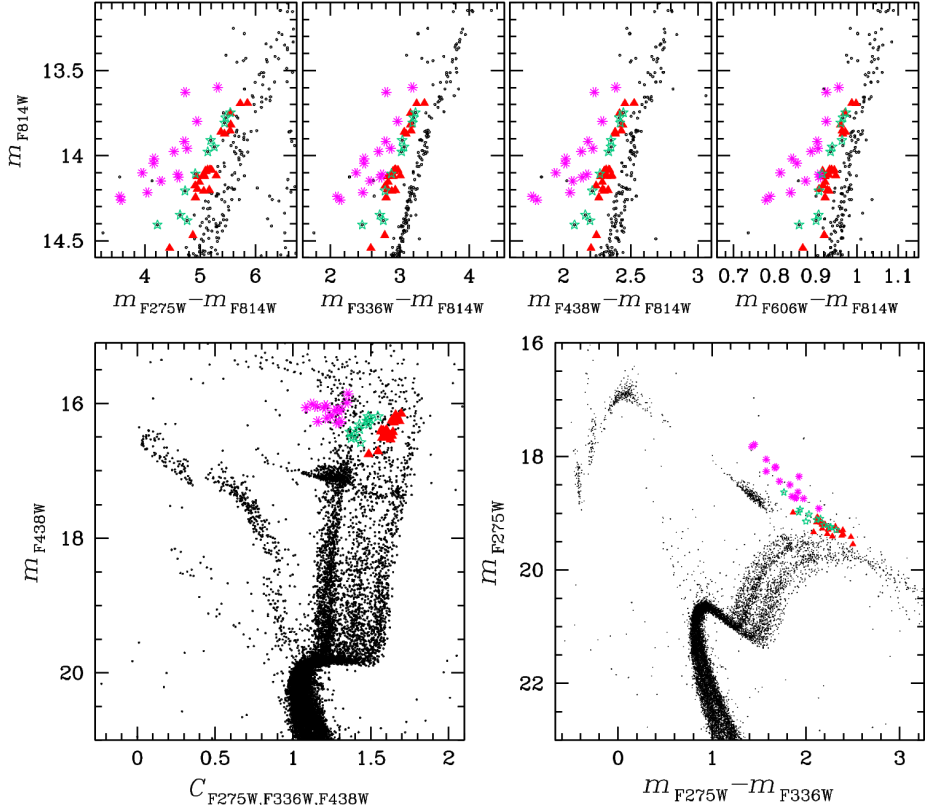}
      \caption{\textit{Upper panels:} $m_{\rm F814W}$ vs.\,$m_{\rm X}-m_{\rm F814W}$ CMDs of AGB and RGB stars in NGC\,2808 ($X$=F275W, F336W, F438W, F606W). \textit{Lower panels:} $m_{\rm F438W}$ vs.\,$C_{\rm F275W, F336W, F438W}$ (left) and $m_{\rm F275W}$ vs.\,$m_{\rm F275W}-m_{\rm F336W}$ CMD (right). Red, aqua, and magenta symbols represent the three groups of  $AGB_{\rm I}$,  $AGB_{\rm II}$, and  $AGB_{\rm III}$ stars defined in the lower-left panel.}
          \label{NGC2808agb}
   \end{figure}

Figure~\ref{NGC2808agb} illustrates our analysis of the AGB of NGC\,2808. The upper panels show a collection of $m_{\rm F814W}$ vs.\,$m_{\rm X}-m_{\rm F814W}$ CMDs zoomed around the AGB ($X$=F275W, F336W, F438W, F606W). The 51 stars with $m_{\rm F814W}>13.5$ that, on the basis of their position in these CMDs, are probable AGBs have been marked with colored symbols. 

The lower-left panel shows the $m_{\rm F438W}$ vs.\,$C_{\rm F275W, F336W, F438W}$ diagram, where AGB stars are distributed along three distinct sequences. This feature is a signature of multiple stellar populations along the AGB. We defined three groups of $AGB_{\rm I}$, $AGB_{\rm II}$, and $AGB_{\rm III}$ stars that include 25, 11, and 15 stars,  colored red, aqua, and magenta, respectively. These colors are used consistently in Fig.~\ref{NGC2808agb}.
 $AGB_{\rm III}$ is bluer than the remaining AGB stars in all the CMDs, while $AGB_{\rm II}$ are slightly bluer than $AGB_{\rm I}$ in all the CMDs apart from the $m_{\rm F814W}$ vs.\,$m_{\rm F336W}-m_{\rm F814W}$ CMD where these two groups of AGB stars are almost overlapping.
The $m_{\rm F275W}$ vs.\,$m_{\rm F275W}-m_{\rm F336W}$ CMD in the lower-right panel of Fig.~\ref{NGC2808agb} shows that $AGB_{\rm III}$ are brighter than the other AGB stars in F275W and that $AGB_{\rm II}$ are, on average, brighter than $AGB_{\rm I}$.

Theoretical models predict that hot HB stars would undergo a transition to an extended blueward nose excursion and exhibit bluer colors than the progeny of cold HB stars  when reaching the AGB (e.g.\,Gingold\,1976).
It is tempting to speculate that the group of $AGB_{\rm III}$ stars is the progeny of helium-rich HB, $AGB_{\rm I}$ stars have primordial helium, and the $AGB_{\rm II}$ belongs to a population with intermediate composition.
 The $AGB_{\rm I}$ stars host 49$\pm$11\% of the total number of AGB stars in agreement, within the large error bar, with the total fraction of the
three helium-poorer RGB-A, RGB-B, RGB-C stars, which include half of the RGB stars of NGC\,2808. The fraction of $AGB_{\rm II}$, and $AGB_{\rm III}$ stars are 22$\pm$6\%, and 29$\pm$8\%, respectively. These numbers only vaguely resemble the fraction of RGB-D ($\sim$31\%) and RGB-E ($\sim$19\%), though we admit there is some arbitrariness in selecting the AGB members of the three groups. 

We conclude that further analysis, possibly based on the synergy between spectroscopy and photometry, is needed to connect the triple AGB with the multiple RGBs of NGC\,2808. Noticeably, stars with extremely thin H-rich envelopes miss the AGB phase, and move towards the white dwarf cooling sequence. 
It is not possible to firmly establish from the present dataset if these `AGB Manque' stars are present in NGC\,2808 or not.

\section{Discussion}
\label{sec:discussion}
 In this paper we have used multi-wavelength {\it HST} photometry to investigate multiple stellar populations in the GC NGC\,2808 as part of the {\it Hubble Space Telescope UV Legacy Survey of Galactic GCs} project (Piotto et al.\,2015).\\
Our basic results can be summarized as follows.

\begin{itemize}
\item  We have identified five distinct stellar groups along the RGB of NGC\,2808, namely A, B, C, D, and E, which contain 5.8$\pm$0.5\%, 17.4$\pm$0.9\%, 26.4$\pm$1.2\%, 31.3$\pm$1.3\%, and 19.1$\pm$1.0\% of the total number of RGB stars with $12.25<m_{\rm F814W}<17.70$, respectively. The five stellar populations have been also detected along the MS althougth the separation between MS-A, MS-B, and MS-C is less evident than in the case of the RGB. We have found that the red MS discovered by Piotto et al.\,(2007) is composed of populations A, B, and C, while their middle and the blue MS correspond to population D and E, respectively.

\item We have exploited high-resolution spectroscopy from literature to infer the abundance of Na, O, Al, and Mg for the five stellar populations. 
 First of all, we have identified the RGB-A--E stars for which chemical abundances from high-resolution spectroscopy are available.
 Specifically, Carretta et al.\,(2006) and Carretta\,(2014) have found large star-to-star variations of  [Na/Fe], [O/Fe], [Al,Fe], and [Mg/Fe] and identified three groups of O-normal, O-poor, and O-super-poor stars.
 We have matched the sample by Carretta and collaborators with our multi-wavelength photometry and found that 32 of their stars belong to populations B, C, D, and E as defined in this paper.  

Using the spectroscopic information, we found that population B has solar sodium-to-iron abundance ratio and is enhanced in oxygen ([O/Fe]$\sim$0.3 dex). Population C is enhanced in sodium ([Na/Fe]$\sim$0.2) dex and slightly depleted in oxygen by $\sim -$0.1 dex with respect to population B.
 Populations D and E are both sodium rich and oxygen poor and have [Na/Fe]$\sim$0.4 and [Na/Fe]$\sim$0.8, and [O/Fe]$\sim-$0.4 and [O/Fe]$\sim-$0.7, respectively. Unfortunately, no population-A RGB  stars have spectroscopic information.
  Abundances of magnesium and aluminum are available only for three population-B and two population-C stars. All of them are distributed around [Mg/Fe]$\sim$0.4 and [Al/Fe]$\sim$0.1 with the two population-C stars being, on average, slightly enhanced in [Al/Fe] by $\sim$0.1 dex. A larger sample is mandatory to establish if such a difference in Al is significant or not. 
 
\item We have inferred the content of helium, C, N, O for the five populations of NGC\,2808. To do this, we have followed the method by Milone et al.\,(2012b) and compared the observed colors with predictions from synthetic spectra. 
 We found that populations D and E are enhanced in helium by $\sim$0.06 and $\sim$0.11, respectively, with respect to the populations B and C which share almost the same helium abundance.  This helium difference follows from the assumption that population A and B have the same metallicity.  From this assumption it also descend that population A has $\sim$0.03 less helium than both population B and C. Planned spectroscopic observations in particular of population A stars may allow us to test this assumption, and indicate whether the photometric differences between these two populations can be ascribed to iron and oxygen differences, rather than to helium. If we assume that population A has primordial helium (Y=0.246), our results indicate that population-E stars are highly helium enhanced up to $Y \sim$0.39.

 Note that it is the effect of helium-abundance variations on stellar temperatures that mostly causes optical and UV colors to change.  Indeed stars with the same luminosity but different $Y$ have different effective temperature and gravity. On the contrary, helium has a marginal effect on the stellar atmosphere. (Sbordone et al.\,2011).

The comparison of observed and synthetic colors allow us to also estimate the average abundance of C, N, and O for each stellar population. We found that populations C, D, and E are enhanced in nitrogen by $\sim$0.5, $\sim$0.6, and $\sim$0.8 dex with respect to population B, while population A has slight lower nitrogen ($\Delta$[N/Fe]$\sim$0.1) than population B. Both oxygen and carbon anti-correlate with nitrogen.

\item
 We have detected the RGB bump of populations B, C, D, and E. 
 In visual filters, which are marginally affected by light element-variations, the bump of population B and C have almost the same luminosity, while the RGB bump of population D and E are $\sim$0.07 and $\sim$0.17 mag brighter. 
 The comparison of the observed bump luminosity with theoretical models suggests that populations D and E are more helium rich than both population B and C by $\sim$0.03 and $\sim$0.06 dex, respectively. 

\item
We confirm that the HB of NGC\,2808 is multi-modal with four main HB segments.
 In addition, we confirm that the red HB is inconsistent with a simple stellar population as suggested by the fact that the two groups of sodium-rich and sodium-poor stars  identified by Marino et al.\,(2014) populate different regions along the red HB.
  
\item
 The AGB hosts three main sequences that are well distinguishable in the $m_{\rm F438W}$ vs.\,$C_{\rm F275W,F336W,F438W}$ diagram. This finding indicates that the AGB of NGC\,2808 hosts multiple stellar populations.

\end{itemize}

In conclusion,
the most astonishing property of this cluster is certainly its extremely complex stellar populations, as illustrated in Figs.~\ref{SummaryNGC2808} and \ref{seleRGBs}.
  We have distinguished five discrete stellar populations, but a closer look at the Hess diagram shown in Fig.~\ref{seleRGBs}
suggests that reality may be even more complex. Indeed, all five clumps  appear to have some internal structure, as if each of them could further split in two components, or having an internal spread in its photometric properties which may result from a small spread in chemical composition. This is particularly evident for populations B and C. We notice that our photometric data have an accuracy of $\sim 0.01$ mag, sufficient to resolve individual clumps, but distinguishing between an internal spread or a multiplicity would require larger samples of stars.

The clear {\it discreteness} of the five (main) populations is a fact that every scenario for the formation of GCs must be able to account for. As emphasized in  Paper\,I, this kind of discreteness is indeed an ubiquitous property among GCs
and can not be ignored. It suggests that star formation occurred in a sequence of discrete events interleaved by periods of inactivity while the chemical composition of the interstellar medium was changing. However, we postpone to a future paper of this collaboration a dedicated discussion as to whether the various proposed scenarios can comply with this and the other observational constraints
illustrated in Paper\,I.

It also goes beyond the scope of this paper to try to identify in which temporal sequence the various multiple populations may have been generated, but we need to at least try to identify the {\it first} generation of this cluster.  Population B is oxygen rich and sodium poor (see Fig.~8) and also is rather populous ($\sim 17\%$ of the total in our sample), so it is the obvious candidate for being the first generation. However, if this is the case, we have a problem with population A,  which is redder than population B in both $m_{\rm F275W}-m_{\rm F814W}$ and $m_{\rm F336W}-m_{\rm F438W}$  and for which no spectroscopic abundances are available from Carretta et al.\,(2006) and Carretta\,(2014). We have argued above that if Population B and A would have the same metallicity, than the helium abundance of population A should be lower that that of population B by $\sim 0.03$. Since no physical mechanism is known that could {\it deplete} on such scale  the helium abundance below its Big Bang value, one would be forced to consider population A as the first generation. This is quite unpalatable, as population A represents only $\sim 6\%$ of the whole population of the cluster.  Even if A$+$B together are regarded as the first generation, still they make only $\sim 23\%$ of the whole population sampled by WFC3 at the center of NGC\,2808. However, as in the case of other clusters, the first generation may be less centrally concentrated than subsequent generations and a more extensive mapping of this cluster is required to measure the overall fractions of the various populations.

The alternative is to relax the assumption of these two populations having the same metallicity. Like in other clusters  (e.g. M\,22, NGC\,1851, M\,2, NGC\,5286), a small fraction of the core-collapse supernovae from the first generation may have contaminated the interstellar medium while such stellar population was still in the making. An increase of [Fe/H] by $\sim 0.1-0.2$ dex, associated to a parallel increase in oxygen as expected from core collapse supernovae, could then account for the photometric differences between population A and B. Our group is already engaged in high-resolution spectroscopic observations with GIRAFFE at the Very Large Telescope and spectra of stars in the five RGBs of NGC\,2808 are going to be obtained soon. In particular, the abundances of iron, carbon, nitrogen, oxygen and sodium will be measured for population A, hence assessing whether this interpretation is viable.

 Moreover it should be noticed that, although some authors have suggested that multiple sequences are associated to distinct generations of stars, the possibility that GCs have experienced multiple or prolonged events of star formation is still strongly debated.  We refer to papers by Bastian et al.\,2013; Cabrera Ziri et al.\,2014, 2015; Niederhofer et al.\,2014; D'Ercole et al.\,2008, 2010; D'Antona et al.\,2005; Renzini et al.\,2008; Decressin et al.\,2007; Denissenkov et al.\,2015 and references therein for critical discussion and for various scenarios and interpretations of multiple stellar populations.

\begin{acknowledgements}
We warmly thank David Yong, who has performed the statistical analysis with the Mcluster CRAN package, described in the Sect.~3.1 and Aaron Dotter and Bob Sharp for useful discussions on the statistical significance of the stellar populations. The anonymous referee and the {\it Statistical Editor} of the journal, Prof.\,Eric Feigelson, have provided several suggestion that have improved the quality of the paper.  
APM and HJ acknowledge support by the Australian Research Council through
 Discovery Early Career Researcher Award DE150101816 and Discovery Project grant DP150100862. SC and GP acknowledge partial support by PRIN-INAF 2014. GP acknowledges partial support by Progetto di Ateneo (Universita' di Padova) 2014.

\end{acknowledgements}
\bibliographystyle{aa}

\end{document}